\documentclass[preprint,12pt]{elsarticle}

\usepackage{graphicx}
\usepackage{subcaption}

\usepackage{amssymb}
\usepackage{mathtools}

\newcounter{bla}

\journal{Computer Physics Communications}

\begin{document}

\begin{frontmatter}



\title{picFoam: An OpenFOAM based electrostatic Particle-in-Cell solver}


\author[a]{Christoph K\"uhn}
\author[a]{Rodion Groll}

\cortext[author] {Corresponding author.\\\textit{E-mail address:} christoph.kuehn@zarm.uni-bremen.de}
\address[a]{Center of Applied Space Technology and Microgravity, University of Bremen}

\begin{abstract}
picFoam is a fully kinetic electrostatic Particle-in-Cell(PIC) solver, including Monte Carlo Collisions(MCC), for non-equilibrium plasma research in the open-source framework of OpenFOAM. The solver's modular design, based on the same principles used in OpenFOAM, makes it highly flexible, by allowing the user to choose different methods at run time, and extendable, by building upon templated modular classes. The implementation of the PIC method employing the finite volume method, allows it to simulate on arbitrary geometries in one to three dimensions. OpenFOAM's barycentric particle tracking is used effectively to performe charge and field weighting from the Lagrangian particle based description to the Eulerian field description and backwards without computational expensive particle searching algorithm. picFoam also includes open and general circuit boundary models for the description of real plasma devices. 
\end{abstract}


\end{frontmatter}



{\bf PROGRAM SUMMARY}

\begin{small}
\noindent
{\em Program Title:} picFoam \\
{\em Developer's repository link:} https://github.com/TFDzarm/picFoam \\
{\em Licensing provisions:} GPLv3 \\
{\em Programming language:} C++ \\
\end{small}

\section{Introduction}

The Particle-in-Cell(PIC) method is a numerical technique combining the Lagrangian description of a fluid with the Eulerian description on a mesh. The method can be applied to a wide range of problems, one of its more popular application lies in the simulation of plasmas.
In its use as a research tool for plasmas, the motion of charged particles, electrons and ions, is described in a Lagrangian manner, while electromagnetic fields generated by the particles are calculated on the computational mesh, from where these fields influence the motion of particles.\\
The method has its roots set in the late 1950s with the self-consistent calculations of Buneman and Dawson. They simulated up to 1000 particles including interactions between them by solving Coulomb's law directly. Over the years improvements were made. In this respect, one important step was the introduction of particle-mesh methods. In these the Poisson equation is solved on a computational mesh, this greatly increased the number of particles which can be simulated. In the 1970s the PIC scheme was formally defined, and a decade later the classic texts by Birdsall and Langdon, and Hockney and Eastwood were published that remain important references to this day\cite{Verboncoeur05}. Since then the growing plasma research community has continued to improve the method, with the inclusion of electromagnetic schemes, introduction of boundary treatments, including external circuits, which allows modeling of real plasma device, and a steadily refinement of particle collision methods. Today the Particle-in-Cell method represents a powerful tool regarding plasma research which is used on par with experiments to gain insight into the complex nature of plasmas.\\
In this paper we describe a new code, called picFoam, and its validation implementing the electrostatic PIC method with Monte Carlo Collisions (MCC) in a library extension to OpenFOAM\cite{Weller98} (Open-source Field Operation And Manipulation) an open source numerical toolbox written in C++. OpenFOAM is primarily designed as numeric library for problems of computational fluid dynamics (CFD), implementing the finite volume method and has high performance parallel computing capabilities using MPI. It is highly flexible and extensible offering a wide range of libraries, solvers, pre- and post-processing tools. Among the standard solvers OpenFOAM offers particle simulation techniques including the Direct Simulation Monte Carlo (DSMC) and molecular dynamics (MD) methods building upon a robust particle tracking algorithm. Based on these particle tracking algorithms, picFoam extends OpenFOAM's capabilities in the area of plasma research. In contrast to other modern open source PIC software applications like XOOPIC\cite{Xoopic95}, EPOCH\cite{Arber15}, SMILEI\cite{Smilei18} (to name a few), picFoam, at the moment, only supports the solution of electrostatic problems in combination with a stationary magnetic field. However, by embedding the new solver into the OpenFOAM framework picFoam gains the ability to interact with OpenFOAM's powerful tool sets, including the pre- and post-processing tools and meshing capabilities. Moreover, the implementation of the PIC method using the finite volume method allows for the use of arbitrary meshes in contrast to the other implementations mentioned before using the finite differences method. As noted by Averkin et al.\cite{Averkin18} there are few implementations of the PIC scheme on unstructured meshes. In their work they present an electrostatic code implemented using a finite volume formulation and support for the simulation of bounded plasmas, which they used to simulate the plasma flow around a small satellite in low earth orbit. Moreover, in their code conductors can be driven by an RLC circuit as in picFoam, however, the code is not openly available. In contrast PICLas\cite{Fasoulas19} is an open source code combining the collisionless electromagnetic PIC method and DSMC method for neutral reactive flows. It is implemented using a discontinuous Galerkin spectral element method and has a wide range of applications ranging form laser plasma interaction to the simulation of re-entry vehicles. pdFoam developed by Capon et al.\cite{Capon2017} is an OpenFOAM based hybrid electrostatic PIC solver, which is used to study the interaction of near earth objects and space environment. The code operates by approximating the electron particle distribution as fluid using the Boltzmann relation but can also handle fully kinetic simulation. However, the code does not handle elastic and inelastic collisions between electrons and neutral particles, Coulomb collisions, and lacks circuit boundary conditions.

\begin{figure}
	\centering
	\begin{subfigure}{.49\textwidth}
		\includegraphics[width=.99\textwidth]{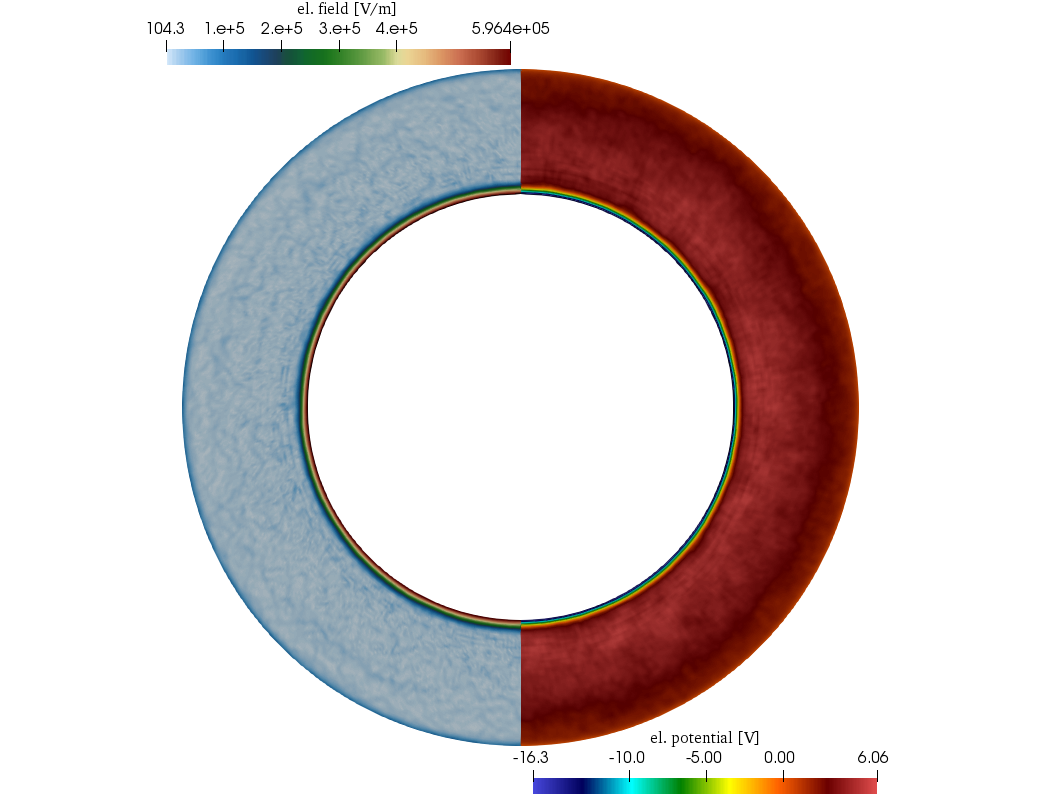}
		\caption{Electric discharge in an annular gap. The inner ring represents the cathode emitting electrons at a constant current density and the outer ring represents the grounded anode.The left half is showing the electric field and the right half is showing the electric potential.}
		\label{fig:Discharge}
	\end{subfigure}
	\begin{subfigure}{.49\textwidth}
		\includegraphics[width=0.99\textwidth]{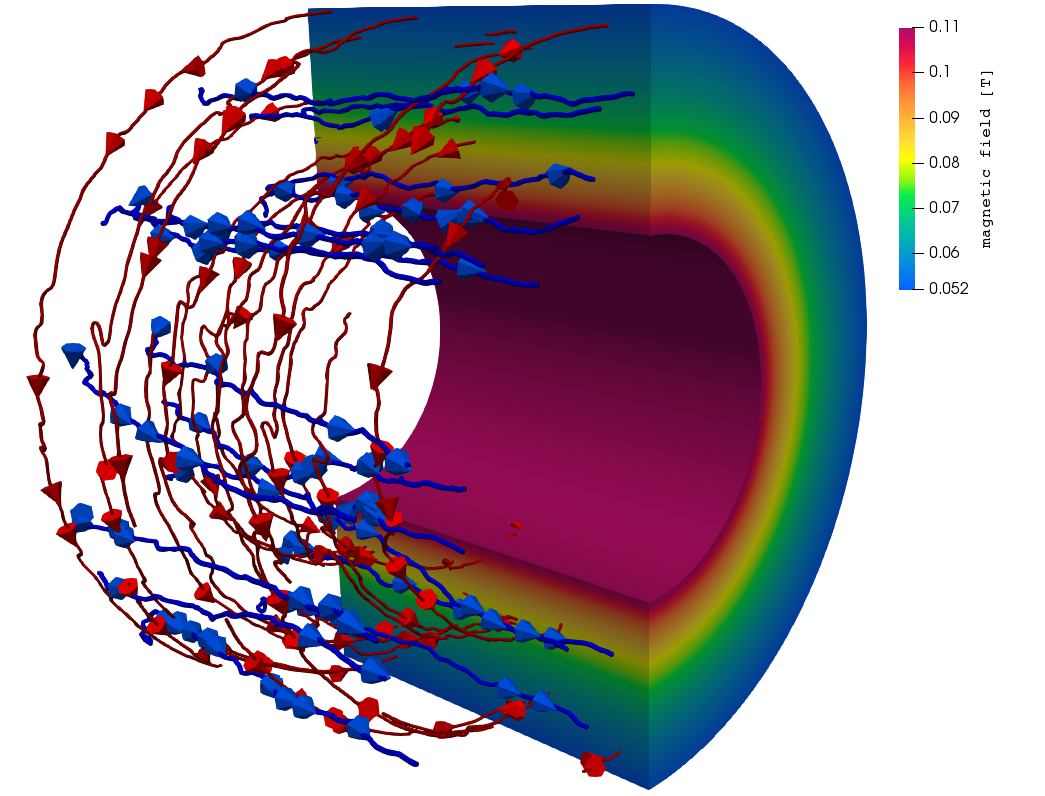}.
		\caption{Hall effect in an annular channel, streamlines depict the mean velocity of species. Heavy ions (blue) move in axial direction, accelerated by the electric field. Electrons (red) move in azimuthal direction accelerated by the  $\mathbf{E}\times\mathbf{B}$-drift.}
		\label{fig:HallEffect}
	\end{subfigure}
	\label{fig:picFoamExamples}
	\caption{Simulations on unstructured meshes performed by picFoam.}
\end{figure}

In contrast, picFoam's intention is to be applicable to a wide range of electrostatic problems. With the inclusion of models for bounded plasmas such as circuit boundary conditions picFoam is capable of simulating real plasma devices. In combination with MCC schemes incorporating inelastic collision events like excitation and ionization through electron-impact, the behavior of plasmas inside electric propulsion systems can be simulated. For instance Fig.\ref{fig:Discharge} shows simplified simulation of a discharge between an inner cathode and outer anode in an annular gap. Fig.\ref{fig:HallEffect} shows the movement of ions and electrons in the discharge chamber of a Hall thruster geometry. In following sections we present the code and validations for all submodules included in picFoam. The software's current release can be obtained from its public repository\cite{github-picFoam} where it has been released under the same general public license as OpenFOAM. 

\section{Software description}

\subsection{Overview of the implementation}

picFoam implements the electrostatic Particle-in-Cell method in a library extension to OpenFOAM, thus the governing Poisson equation Eq.\ref{eq:Poisson} is solved by using the finite volume method to obtain the electric field $\mathbf{E}$, which is then used to accelerate the simulated particles. In addition the solver can incorporate a constant magnetic field $\mathbf{B}$ to the particle's integration of motion (see section \ref{sec:ParticlePusher}).

\begin{equation}
\nabla^2 \phi = -\frac{\rho_c}{\epsilon_0}\textnormal{,\hspace*{0.5cm}}\mathbf{E} = -\nabla\phi
\label{eq:Poisson}
\end{equation}

\begin{figure}[h]
	\centering
	\includegraphics[width=0.8\textwidth]{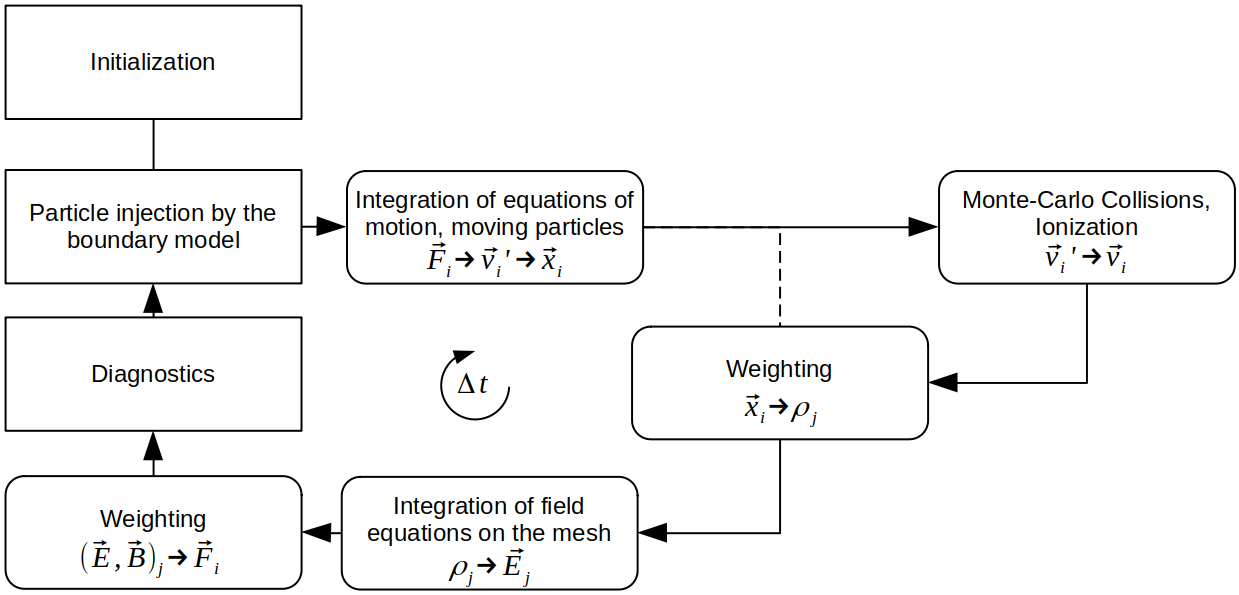}
	\caption{Particle-in-Cell scheme. The classic computation cycle\cite{BirdsallLangdon05} is depicted using boxes with rounded off edges, additional boxes show further steps performed by picFoam.}
	\label{fig:PICscheme}
\end{figure}

\noindent Simulations can be run in one to three spatial dimensions using arbitrary structured meshes. Particles are described with three velocity components thereby placing them in variable x plus 3 dimensional phase space (xD3V). One key aspect of the PIC method is that simulated particles include a number of real particles $\omega_p$, within picFoam this number is arbitrary and defined on a particle to particle basis.\\
In general, simulations conducted with picFoam follow the PIC-MCC scheme depicted in Fig.\ref{fig:PICscheme}. Here the classic scheme\cite{BirdsallLangdon05} including Monte Carlo Collisions is drawn using boxes with rounded off edges, additional steps taken by picFoam during a time step are drawn using boxes with sharp edges.\\

The simulation process can be summarized as follow. First, preceding a simulation, an initialization step is performed by using the application \textbf{picInitialise}, which is provided by the picFoam repository. With this pre-processing tool particles are distributed in the domain, an initial electric field $\mathbf{E}$ is solved, and models, selected by the user, are initialized as described in section \ref{sec:Initialization}.\\
After the initialization, the simulation is run using the application picFoam. Every time step the scheme processes the following steps until the end time is reached.
\begin{enumerate}
	\item[] \textbf{Step 1}: Injection of new particles from boundaries. picFoam supports various boundary and emission models as will be discussed in section \ref{sec:BoundaryModels}.
	\item[] \textbf{Step 2.1}: Integration of the velocities of all particles using the leapfrog scheme and the various integration algorithms (see section \ref{sec:ParticlePusher}).
	\item[] \textbf{Step 2.2}: Moving all particles with their updated velocities using OpenFOAM's barycentric particle tracking  (see section \ref{sec:OFtracking}).
	\item[] \textbf{Step 3}: Performing Monte Carlo Collisions, this includes models for binary collisions between neutral species and Coulomb collisions of charged particles (see section \ref{sec:collision}).
	\item[] \textbf{Step 4}: Weighting the charges of all particles to the mesh as will be described in section \ref{sec:ParticleWeighting}.
	\item[] \textbf{Step 5}: Integration of the field equation Eq.\ref{eq:Poisson} using the finite volume method and OpenFOAM's matrix solvers.
	\item[] \textbf{Step 6}: Weighting the electric field to the particle position (see section \ref{sec:ParticleWeighting}), which in turn is used to integrate the equation of motion in step 2.1.
	\item[] \textbf{Step 7}: Calculation of diagnostics as will be described in section \ref{sec:Diagnistics}.
\end{enumerate}

\noindent Each procedural step is implemented in a modular way using submodule classes, which rely heavily on the template programming techniques used by OpenFOAM and allows for the selection of the various models at run time. 

\begin{figure}[h!]
	\centering
	\includegraphics[width=0.3\textwidth]{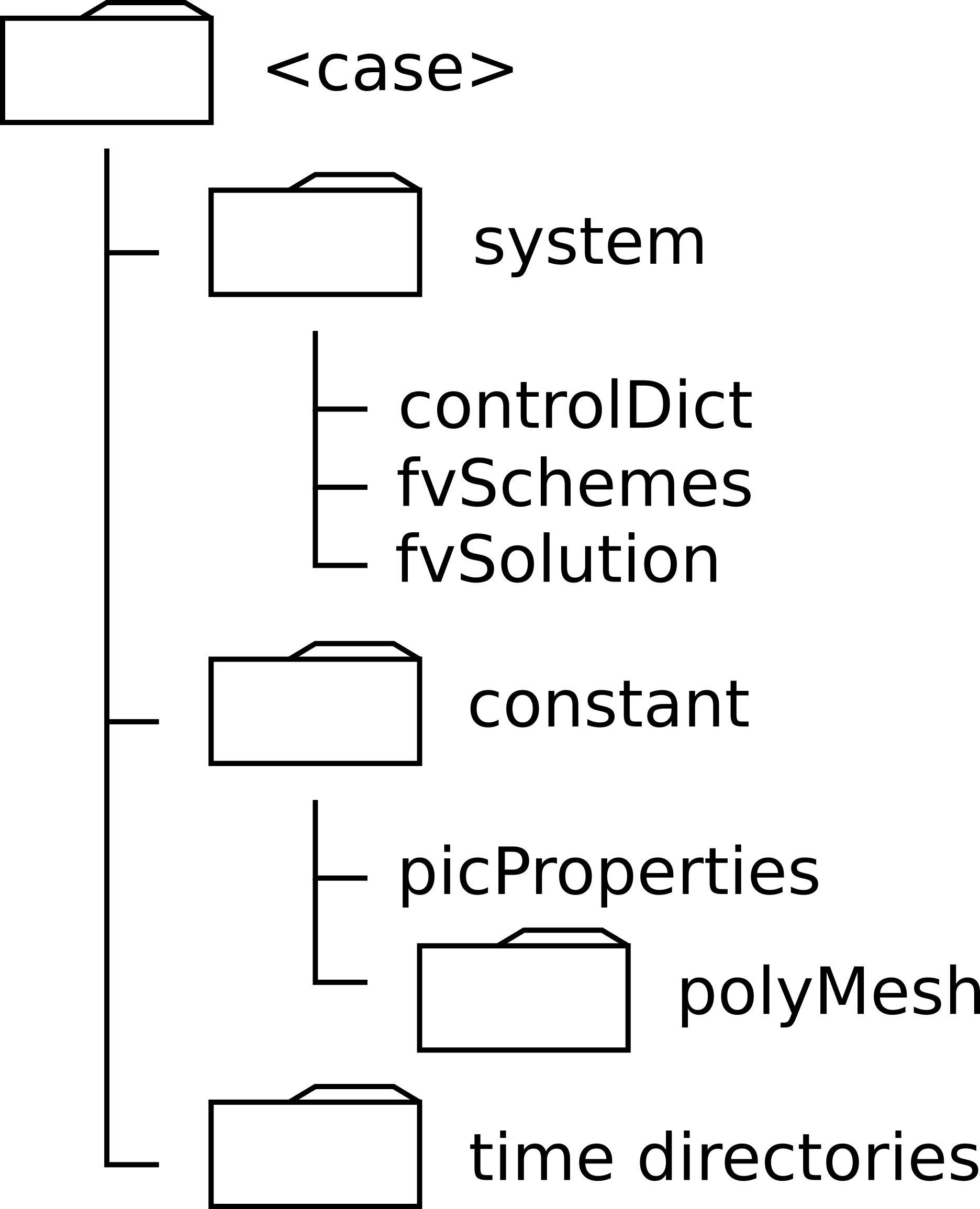}
	\caption{Schematic structure of a picFoam simulation case.}
	\label{fig:OFstructure}
\end{figure}

\subsection{Case structure}

Simulations conducted with picFoam are set up using OpenFOAM case structure consisting of time directories, a constant, and a system directory. The time directories contain the fields calculated by the solver at different time points. Simulations start from initial fields in the "0" time directory, here boundary conditions for the electrostatic potential field \textbf{phiE} need to be supplied by the user as well.\\
The constant directory containing the numerical mesh in a subdirectory called polyMesh and constant settings concerning the simulation in the dictionary file \textbf{picProperties}. Meshes can be generated with OpenFOAM's meshing utilities foremost blockMesh and snappyHexMesh, however, the importing of several popular mesh formats from other meshing utilities and conversion to OpenFOAM's own format is also supported.\\
In the \textbf{picProperties} dictionary the user sets physical properties of the considered species and selects models for all submodules, for more details the reader is referred to the corresponding sections of this paper.\\
The system directory includes parameter settings for the solution procedure, within this directory three files are required for the functionality of OpenFOAM: the \textbf{controlDict}, which controls time and solution output, \textbf{fvSchemes}, which controls the schemes used in discretization process, and \textbf{fvSolution}, which determines the matrix solver used. Additional files for controlling the utilities provided by OpenFOAM, e.g. the decomposition in parallel processing are also supplied in the system directory.

\subsection{Initialization}
\label{sec:Initialization}
To initialize a simulation case the picFoam repository includes picInitialise, a pre-processing tool used to distribute particles in the domain and initialize the submodules of the solver. It is build on the same principle as its main application using submodules to make it easy for a user to extend its features.\\
For the initialization process the user supplies macroscopic quantities like temperatures and number densities for each species, and picks an initialization model in the \textbf{[case]/system/picInitialiseDict} dictionary. By default picInitialise provides a number of different initialization models. These models include, listing only a few, an equipartioned distribution of particles from Maxwell-Boltzmann distribution, a quasi-neutral initialization of plasma, an initialization from a list, where positions and velocities of single particles can be supplied, or a model for a uniform distribution of particles. If not required by the initialization model for the distribution of particles, picInitialise automatically calculates the number densities and temperatures for all species.\\
After the distribution of the particles, the application initializes the collision models used by picFoam (see section \ref{sec:collision}) and solves the initial electric field, which is used to prepare the particles for the leapfrog scheme (see section \ref{sec:ParticlePusher}).\\
In the final step the application performs initial checks on restrictions set by the PIC method. These include a special restriction, as shown by C.K. Birdsall \cite{Birdsall91}, where the cell dimensions need to be in the order of the Debye length

\begin{equation}
\lambda_D = \sqrt{\frac{\epsilon_0 k_B /e^2}{n_e / T_e + \sum_j Z_j^2 n_j/T_i}}\,\textnormal{,}
\label{eq:DebyeLength}
\end{equation}

\noindent to suppress the growth of nonphysical instability. In this context the Debye length describes a shielding effect of the electrostatic potential of a charge trough the self arranging of charges with different sign around the particle considered\cite{Bittencourt04}. Here $\epsilon_0 = 8.854... \cdot 10^{-12}$ As/Vm is the vacuum permittivity, $k_B = 1.380... \cdot 10^{-23}$ J/K is Boltzmann's constant, $e = 1.602... \cdot 10^{-19}$ As is the elementary charge, $Z$ indicates the charge number of the ionic species $j$, $n$ represents the number densities and $T$ are the species temperatures. The second restriction owed to the particle pusher, is different for all pushers, however a general stability criterion linked to the explicit leapfrog scheme, used in picFoam, can be formulated \cite{Tskhakaya04}. Revealing that the time step has to be smaller than twice the plasma frequency

\begin{equation}
\omega_{pe} = \sqrt{\frac{n_e\,e^2}{\epsilon_0 m_e}} \textnormal{.}
\label{eq:PlasmaFrequency}
\end{equation}

\noindent The parameter in Eq.\ref{eq:PlasmaFrequency} have the same meaning as before, additionally, $m_e = 9.109...\cdot 10^{-31}$ kg is the rest mass of an electron. The plasma frequency describes the oscillation of electrons around positively charged ions, due to the attractive Coulomb forces and the inertia of the electrons, causing them to pass the equilibrium position\cite{Bittencourt04}. Further monitoring of plasma frequency and Debye length during simulation with picFoam can be turned on in the dictionary \textbf{[case]/constant/picProperties}.\\
For statistical accuracy a high enough number of particles per cell is needed, to ensure this, picInitialise prints statistics on the cell occupancy along with other statistics like particle velocities, number densities and temperatures for all species, after the initialization.

\subsection{Particle pusher}
\label{sec:ParticlePusher}

\begin{figure}
	\centering
	\includegraphics[width=0.4\textwidth]{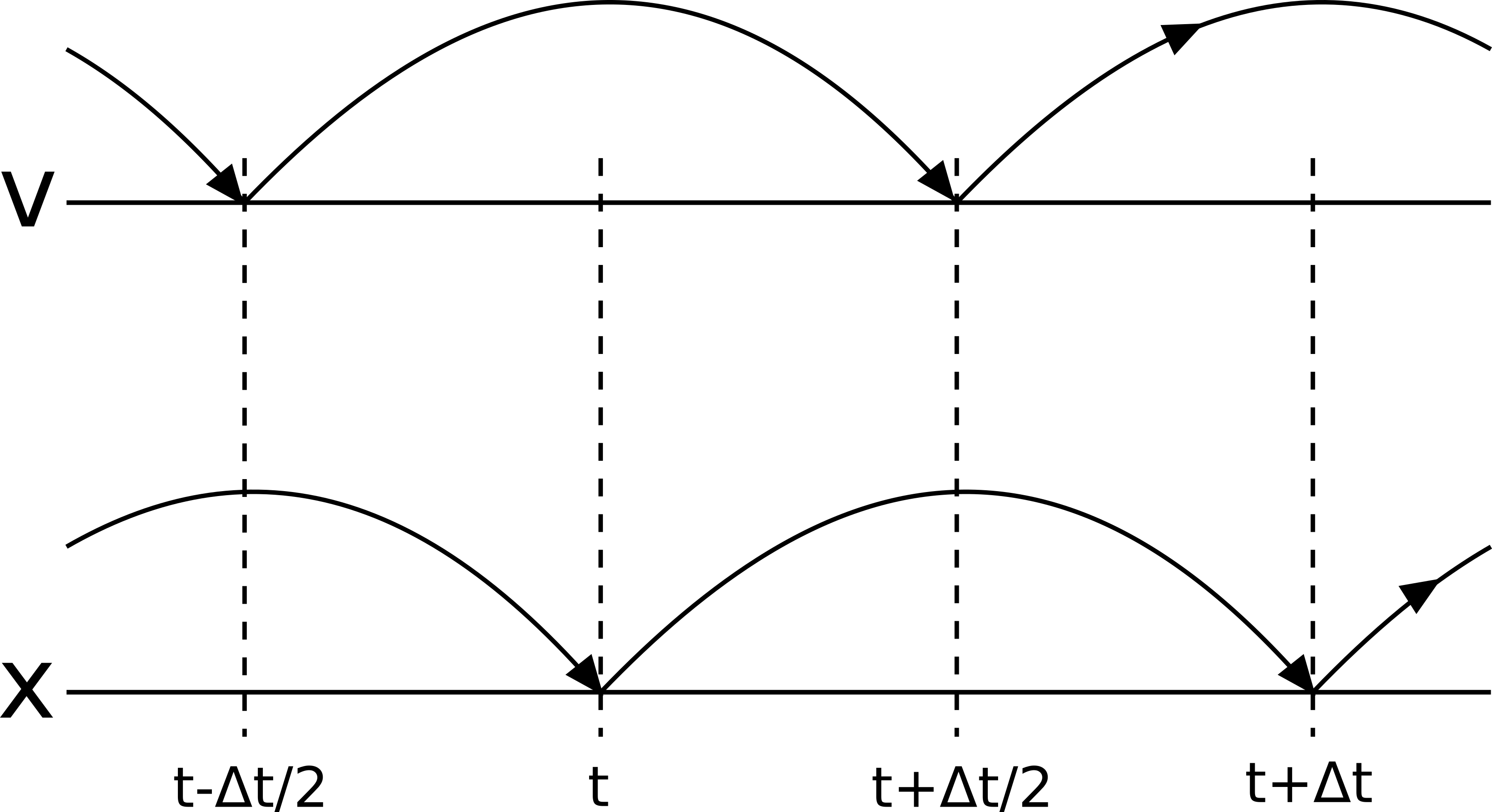}
	\caption{Leapfrog scheme. Initially the velocities of the particles are set back by half a time step, resulting in the temporal leap frogging of spatial and velocity data.}
	\label{fig:leapfrogScheme}
\end{figure}

Commonly in the PIC method the leapfrog scheme\cite{Tajima04} is used for the integration of the particle's equation of motion. In this scheme the average velocity existing at time $t^{n+1/2}$ is used to move the particle forward. To achieve this, the initial velocity of the particles is set back by half a time step during the initialization. Doing so the velocity and position of a particle are not known at the same time, but leap over each other during simulation (see Fig.\ref{fig:leapfrogScheme}).\\
To perform the velocity integration on every particle in the electromagnetic field (picFoam supports the use of a constant magnetic field), the PIC method applies the Lorentz force $\mathbf{F_L}$ to compute a new velocity for each particle.

\begin{equation}
m \frac{d\gamma\mathbf{u}}{dt} = \mathbf{F_L} = q (\mathbf{E}+\mathbf{u}\times \mathbf{B})
\label{eq:VelocityIntegration}
\end{equation}

\noindent In Eq.\ref{eq:VelocityIntegration} the vectors for the electric $\mathbf{E}$ and magnetic $\mathbf{B}$ field are known at integral time steps, while the velocity $\mathbf{u}$ is known at half integer time steps. In this equation the parameter $\gamma = \sqrt{1+\left|\mathbf{u}\right|^2/c^2}$ represents the Lorentz factor.\\
To solve Eq.\ref{eq:VelocityIntegration} various models are implemented, these include the relativistic models of Vay\cite{Vay08}, Higuera and Cary\cite{Higuera17}, and the commonly used model of Boris\cite{Boris70} in non-relativistic and relativistic form. The model is chosen by setting the \textbf{ParticlePusher} entry in the file \textbf{[case]/constant/picProperties}.

\subsubsection{Barycentric particle tracking}
\label{sec:OFtracking}
To move the particles through the domain, picFoam uses OpenFOAM's barycentric particle tracking. This allows for tracking which is defined in terms of displacement, without any search or correction algorithm.
It leads to advantages in the linear weighting algorithm (see section \ref{sec:ParticleWeighting}), and the boundary treatment. In the following paragraph we explain the fundamental steps of this algorithm for a stationary mesh.

\begin{figure}
	\centering
	\includegraphics[width=0.4\textwidth]{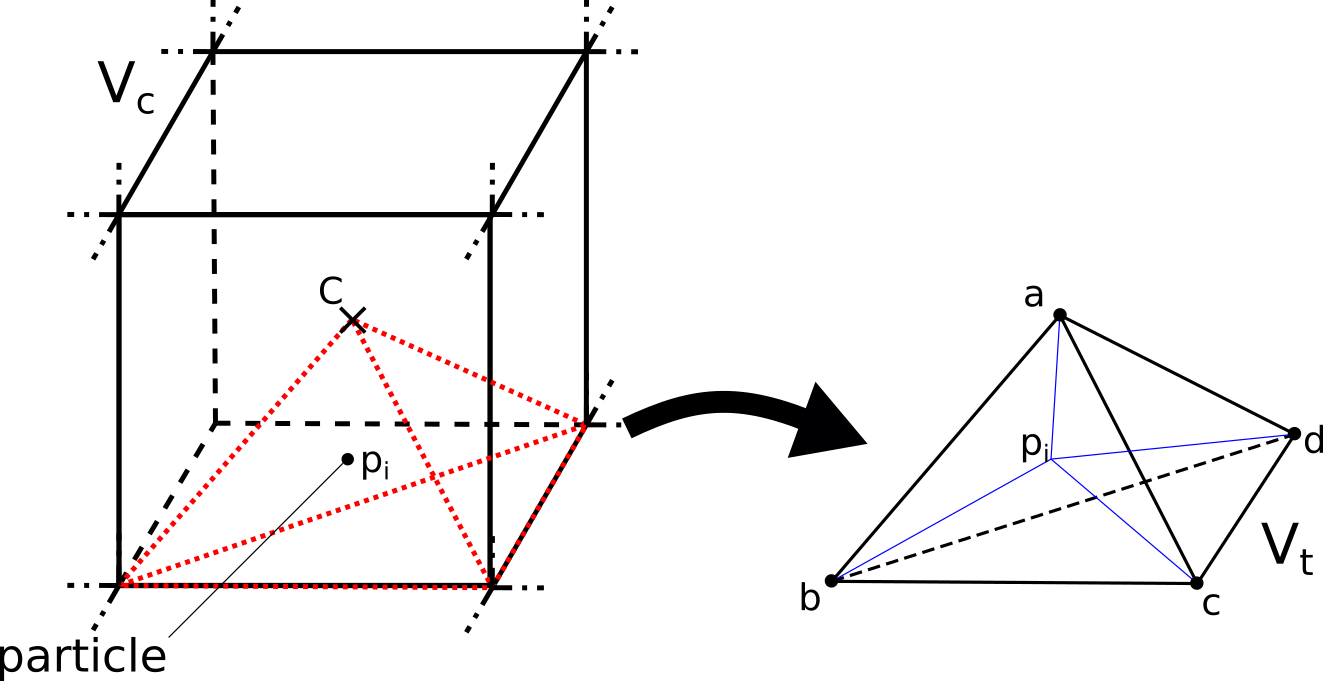}.
	\caption{Illustration of one tetrahedron in the cell decomposition of hexagonal cell and of the barycentric description of a particles position. }
	\label{fig:tetDecompose}
\end{figure}

\noindent Each arbitrarily polyhedral cell in an unstructured finite volume mesh can be decomposed into tetrahedra, these are constructed by triangles on the cell's faces and the cell's center point (see Fig.\ref{fig:tetDecompose}). The tetrahedra are then used to define the position of each particle in terms of topology of the tetrahedron and the place of the particle inside it, specified by barycentric coordinates. These coordinates $\mathbf{y}$ define the Cartesian position $\mathbf{x}$ as a weight of the sum of the tetrahedron's vertices $\mathbf{a},\mathbf{b},\mathbf{c},\mathbf{d}$

\begin{equation}
\mathbf{x} = \mathbf{a} y_1 + \mathbf{b} y_2 + \mathbf{c} y_3 + \mathbf{d} y_4\textnormal{.}
\end{equation}

\noindent Transformation from the Cartesian to the barycentric coordinates are performed using the transformation tensor $\mathbf{\overline{A}}$ constructed from the tetrahedron's edges

\begin{equation}
\mathbf{\overline{A}}\cdot\mathbf{y} = \mathbf{x} - \mathbf{a}\textnormal{,}
\end{equation}

\noindent where the tensor

\begin{equation}
\overline{\mathbf{A}} = \begin{matrix}
[\overrightarrow{ab},\overrightarrow{ac},\overrightarrow{ad}]
\end{matrix}\textnormal{.}
\end{equation}

\noindent A search of the particle's position in the mesh and calculation of the barycentric coordinates is only required once when introducing a new particle into the domain, as barycentric coordinates are handled as primary data and are automatically updated while tracking the particle.\\
The tracking works by moving from tetrahedron to tetrahedron, interacting with all cell faces hit on the way. The linear change of a particle position in Cartesian coordinates is found using

\begin{equation}
\label{eq:CartesianTrack}
\mathbf{x}^{new} = \mathbf{x} + \lambda\mathbf{x}_d\textnormal{,}
\end{equation}

\noindent where $\textbf{x}_d=(1-f)\textbf{u}\Delta t$ is the particle's global displacement, with $f$ as the step fraction keeping track of the already covered global track and $\lambda$ the fraction of the current track to the next face. For more details on the Cartesian particle tracking in older OpenFOAM versions correspond to Macpherson et al.\cite{Macpherson08}.\\
Using the inverse transformation tensor

\begin{equation}
\overline{\mathbf{T}} = \det(\overline{\mathbf{A}}) \overline{\mathbf{A}}^{-1}
\end{equation}

\noindent and

\begin{equation}
\mu = \frac{\lambda}{\det(\overline{\mathbf{A}})}
\end{equation}

\noindent Eq.\ref{eq:CartesianTrack} is transferred to an update of the barycentric coordinates of the particle

\begin{equation}
\label{eq:BarycentricTrack}
\mathbf{y}^{new} = \mathbf{y} + \mu\overline{\mathbf{T}} \cdot \mathbf{x}_d \textnormal{.}
\end{equation}

\noindent Hits with the triangle faces of the tetrahedron are found by setting ($\mathbf{y}^{new})_i = 0$ and solving for the smallest

\begin{equation}
\label{eq:smallestMu}
\mu_i = -\frac{y_{i}}{(\overline{\mathbf{T}} \cdot \mathbf{x}_d)_i} \textnormal{.}
\end{equation}

\noindent Doing this means that there is no division by $\det(\overline{\mathbf{A}})$ anywhere in the algorithm allowing it to function on inverted or degenerate tetrahedra.
The resulting $\mu$ is then used to update the particle's coordinates using Eq.\ref{eq:BarycentricTrack} and the step fraction $f$ using $\det(\overline{\mathbf{A}})\mu$. 
Following this a check if $\mu$ is greater than $\det(\overline{\mathbf{A}})$ reveals if the track of the particle ends in the current tetrahedron. If this is the case, the overall track of the particle is finished, else the hit triangle face of the tetrahedron is checked for a number of cases before continuing the track with the updated step fraction. In the first case, if the triangle does not belong to a cell surface, the particle switches to the adjacent tetrahedron in the current cell and continues its track. If the particle hits a cell surface neighboring another cell, the particle switches to the other cell updating its cell occupancy. If the cell face belongs to an boundary, immediate interaction with the boundary model is possible resulting in a reflection, deletion or more complex interactions.\\
Using this barycentric tracking algorithm the particle's location is always known and the need for a computational expensive mesh searching algorithm after each particle movement is avoided. When compared to other equivalent Cartesian algorithms, the barycentric condition ($\mathbf{y}^{new})_i = 0$ of finding the intersecting triangle on the particle's trajectory is exact, preventing round off errors, associated to the determination of the hit triangle face in other algorithms. This means tracking errors where the particles can get "lost" when hitting the triangles vertices or edges are avoided. For further implementation details of the algorithm correspond to OpenFOAM's software repository \cite{github-OpenFOAM8}.

\subsection{Weighting}
\label{sec:ParticleWeighting}

In order to calculate the forces acting on the particle self-consistently, picFoam has to calculate the charge density, solve the electrostatic Poisson equation and obtain the electric field, which is, in the electrostatic case, the negative gradient of the potential field $\mathbf{E} = -\nabla\phi$. These steps require two interpolations also called weighting. First, the particle's charge is weighted from its position to the numerical mesh and second, the field is weighted back from the electric field, defined on the mesh, to the particle's position \cite{BirdsallLangdon05}. picFoam supports two methods which can be selected at run time in the \textbf{SolverSettings} dictionary in \textbf{[case]/constant/picProperties}.

\subsubsection{Cell average weighting}
\label{sec:CellAverageWeighting}
The simplest method sums the charges $Q = \omega_p q_p$ of every particle $N_p^i$ to their corresponding cell indicated by the superscript $i$ and divides by the cell volume to obtain the charge density field. Since this method directly assigns the charge to each element and divides by the volume, charge is guaranteed to be conserved.

\begin{equation}
\rho_c^i = \frac{1}{V_c^i}\sum_j^{{N_p}^i} Q_j
\end{equation}

\noindent After solving the Poisson equation, the electric field vector of a cell is used on every particle in its corresponding cell to calculate the force acting on it. The cell average method is very fast, since only a summation of the charges is necessary for all cells. The disadvantage of this method are the relatively noisy fields, and therefore high gradients between neighboring cells.

\subsubsection{Volume weighting method}
\label{sec:VolumeWeighting}
The second weighting method implemented in picFoam uses the barycentric coordinates, that are known for all particles at any point in time (see section \ref{sec:OFtracking}). This allows picFoam to effectively calculate the weighting of the charges to surrounding cells. The following paragraph describes the implemented method.\\
Barycentric coordinates are equivalent to the ratios of the sub tetrahedron's volume $V_i$, constructed with position $\mathbf{x}$ of the particle, and the volume $V_{t}$ of the entire tetrahedron the particle is located in\cite{Buning89}. Using these coordinates, charge is weighted to the vertices of the cell using
\begin{equation}
Q^k = \sum_j^{{N_p}^i} Q_j\,\mathbf{c}_k\textnormal{,}
\end{equation}
where $k=2,3,4$ denote the vertices of the tetrahedron that coincide with vertices of the cell. The weighted amount of charge $Q^k$ for $k=1$ is assigned to the cell's center, since the sum of the barycentric coordinates $\sum_i \mathbf{c}_i = 1$, charge is conserved and errors only occur in the order of magnitude of the computational accuracy, as for the cell average method.
In the next step, after communicating the charges on the vertices over processor boundaries and accounting for periodic boundary conditions, the summed charges are distributed back from the vertices to all adjacent cells. Hereby the distance from the vertex to the cell center divided by the sum of the distances to all cell centers, surrounded by the respective vertex, is used. On the cell centered field, the interpolated charges are added to the value $Q^1$ and a subsequent division by the cell volume $V_c$ leads to the charge density $\rho_c^i$, which is used by the field solver.\\
After the new electric field has been computed, it has to be weighted back from the cell-centered field to the particle's position. Hereby the field vectors are interpolated from the cell center to all vertices using the distances to them. Again, after communicating over processor boundaries and accounting for periodic boundary conditions, the barycentric coordinates of the particles are used to linearly interpolate the electric field to the position of the particles.\\
This weighting method has a higher computational cost, when compared to the cell averaging method, but leads to much less noisy fields, which can result in more exact solutions, when compared to the first method (see section \ref{sec:PlasmaOscillation}).

\subsection{Maxwell solver}

Currently, picFoam only supports the solution of the electrostatic Poisson equation Eq.\ref{eq:Poisson}. Like for every other submodule in picFoam, the selection of the model can be done at run time setting the \textbf{MaxwellSolver} entry in the \textbf{[case]/constant/picProperties} dictionary, hereby giving the user the option to choose between no solver and the electrostatic solver. Settings for the discretization process of the finite volume method and the matrix solver are specified in the \textbf{[case]/system/fvSchemes} and \textbf{[case]/system/fvSolution} directories respectively.

\subsection{Boundary models}
\label{sec:BoundaryModels}

Within picFoam there is a variety of boundary models to choose from. Their selection is done in the \textbf{[case]/constant/picProperties} sub dictionary \textbf{BoundaryModels}.\\
There are two base models, which have to be selected, the first specifies the reflective interaction from wall boundaries by setting the \textbf{WallReflectionModel} entry. Here picFoam supports simple specular reflection and more realistic diffusive reflection models dependent on the temperature of the wall. The implementation of these models is borrowed from OpenFOAM's DSMC solver.\\
The second base model defines the boundary condition and interaction with simple patch boundaries, comprised of all boundaries, that are no wall or special boundary like periodic or symmetry boundary conditions. Patches defined by this second base model can represent an open boundary like a free stream boundary implemented similar to the model used in OpenFOAM's DSMC solver or a simple injection model, that emits particles from the boundary according to a supplied injection frequency and velocity model. The models are selected by specifying patch names and the assigned model in the \textbf{PatchBoundaryModels} sub dictionary. In this, picFoam allows for the selection of different boundary models on different patches other than OpenFOAM's DSMC solver. Another class of important boundary models implemented within picFoam are circuit models based on the work of Verboncoeur et al.\cite{Verboncoeur90} (see section \ref{sec:circuitModels}).\\
In addition to the two base models, picFoam also includes so called event models, which can be used additionally to the base models on all boundaries. Their main purpose is to perform diagnostic calculations like deletion statistics, but they also include sputtering models which can be used on e.g. wall boundaries. Patch event models are selected through the \textbf{PatchEventModels} dictionary entry.

\subsubsection{Circuit boundary models} 
\label{sec:circuitModels}

\begin{figure}[h]
	\centering
	\includegraphics[width=0.4\textwidth]{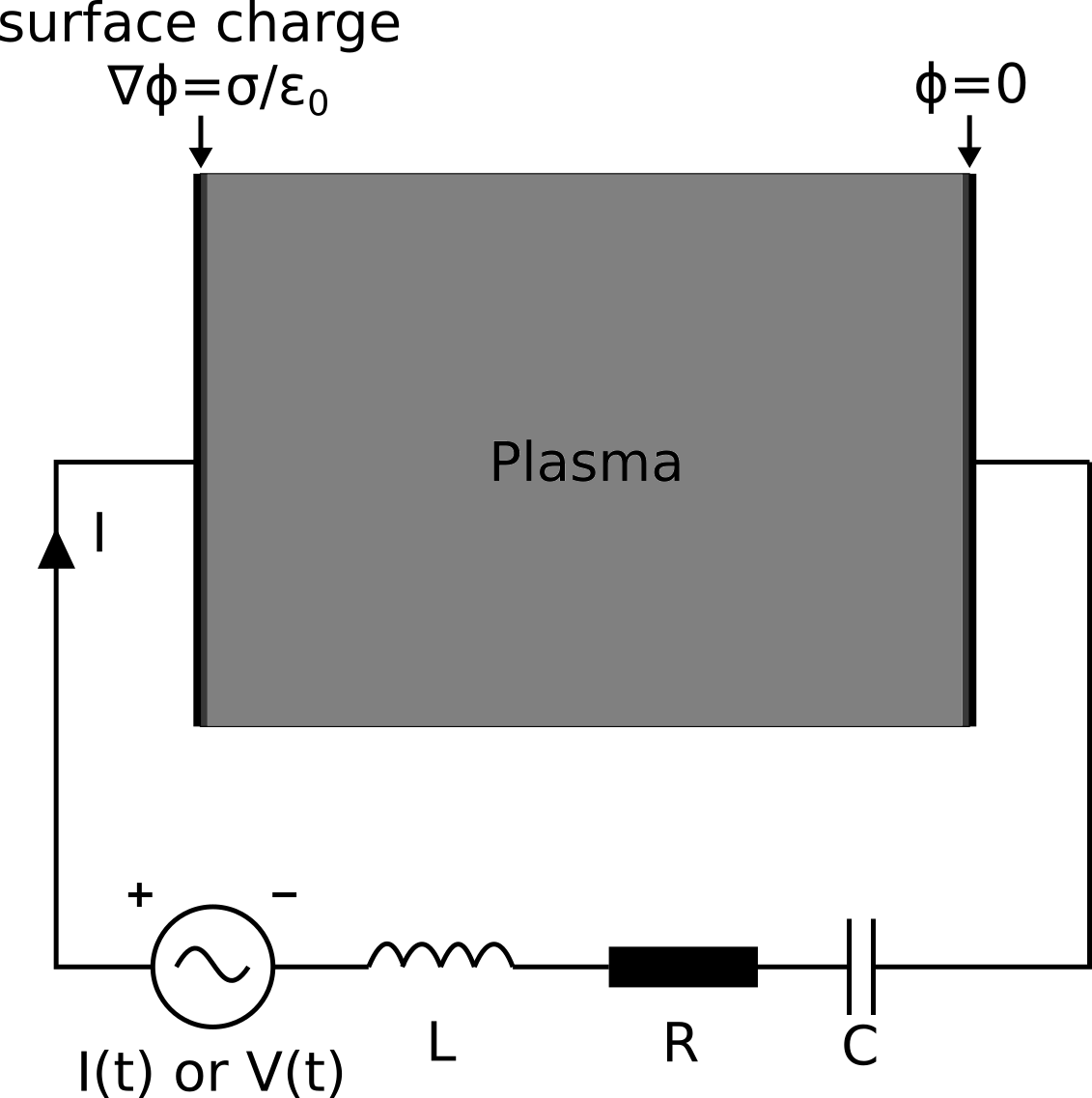}.
	\caption{Scheme of a bounded plasma with external RLC circuit in one dimension.}
	\label{fig:RLCcircuit}
\end{figure}

The addition of circuit boundary models allows picFoam to analyze bounded plasmas. Implemented are simple open circuit and general purpose circuit models consisting of a resistance, an impedance, and a capacity in series (see Fig.\ref{fig:RLCcircuit}). The open circuit models include a floating boundary model, a short circuit ideal voltage, and an ideal current source model.
The usage of these models requires boundary conditions joining the circuit charge to the plasma domain, these are based on Gauss' law Eq.\ref{eq:gausslaw} using the surface charge $\sigma$ and applying it to the boundary surface of finite volume mesh.

\begin{equation}
\label{eq:gausslaw}
\oint_S \epsilon \mathbf{E}\cdot d\mathbf{S} = \int_V\rho_c\, dV + \oint_S \sigma\, dS = Q
\end{equation}

\noindent In electrostatic problems solving the Poisson equation, the relationship

\begin{equation}
\mathbf{E} = -\nabla \phi = \frac{\sigma}{\epsilon}
\label{eq:CircuitBoundaryCondition}
\end{equation}

\noindent is used to specify the surface charge $\sigma$ as a gradient boundary condition for the electrostatic potential. As shown by Vahedi et al.\cite{Vahedi97}, the time variation of the surface charge, may be obtained from Kirchhoff's current loop law

\begin{equation}
\frac{d\sigma}{dt} = {J}_{conv} + \frac{I(t)}{A}\textnormal{,}
\label{eq:SurfaceUpdate}
\end{equation}

\noindent where $I(t)$ is the external circuit current, $A$ the area of the boundary representing the electrode, and $J_{conv}$ is the convective current density supplied to the electrode by charged particles. The discrete finite
differenced form is 

\begin{equation}
A (\sigma^t-\sigma^{t-1}) = Q^t - Q^{t-1} + Q_{conv}^t\textnormal{,}
\label{eq:convCharge}
\end{equation}

\noindent where $Q = \int dt I$ is charge deposited by the external circuit current. The conservation of charge Eq.\ref{eq:convCharge} is used to update the surface charge every time step. Using the general circuit, consisting of a resistance $R$, an impedance $L$, and a capacity $C$ in series, the charge $Q$ existing in the capacitor is advanced using Kirchhoff's voltage law

\begin{equation}
L \frac{d^2 Q}{dt^2} + R \frac{dQ}{dt} + \frac{Q}{C} = V(t) + \Phi_{a} - \Phi_c\textnormal{.}
\label{eq:KirchhoffsVoltageLaw}
\end{equation}

\noindent In the implementation of picFoam, $\Phi_{c}$ is the potential averaged at the driven boundary, while $\Phi_a$ acts as the reference potential for the system and is fixed to zero at the second boundary.\\
Eq.\ref{eq:KirchhoffsVoltageLaw} is differenced using the second-order backward Euler representation as described by Verboncoeur et al.\cite{Verboncoeur90} and transformed into a gradient boundary condition using Eq.\ref{eq:CircuitBoundaryCondition} and Eq.\ref{eq:convCharge}.

\begin{equation}
\nabla \phi = \frac{\phi^t_b-\phi^t_c}{\Delta x} = \frac{1}{\alpha_0 \epsilon_0 A + \Delta x} \Big(-\phi^t_c + \alpha_0 A \sigma^{t-1}
+ \alpha_0( Q_{conv}^t - Q^{t-1}) + (V(t)-K^t)\Big)
\label{eq:RLCBoundaryCondition}
\end{equation}

Eq.\ref{eq:RLCBoundaryCondition} specifies the potential gradient between the boundary $\phi^t_b$ and the value at the cell center $\phi^t_c$ at time $t$. $\Delta x$  is the distance between the cell center and the boundary face and $A$ is the area of the boundary surface, while $\alpha_0$ and $K^t$ have the same definition as in Verboncoeur et al.\cite{Verboncoeur90}.

\begin{equation}
\begin{split}
K^t = \alpha_1 Q^{t-\Delta t}& + \alpha_2 Q^{t-2\Delta t}+\alpha_3 Q^{t-3\Delta t}\alpha_4 Q^{t-4\Delta t}\\
&\alpha_0 = \frac{9}{4} \frac{L}{\Delta t^2}+\frac{3}{2}\frac{R}{\Delta t} +\frac{1}{C}\\
&\alpha_1 = -6 \frac{L}{\Delta t^2}-2 \frac{R}{\Delta t}\\
&\alpha_2 = \frac{11}{2}\frac{L}{\Delta t^2} + \frac{1}{2} \frac{R}{\Delta t} \\
&\alpha_3 = -2 \frac{L}{\Delta t^2}\\
&\alpha_4 = \frac{1}{4}\frac{L}{\Delta t^2}
\end{split}
\label{eq:VerboncoeurParameter}
\end{equation}

When an open circuit is used, the gradient boundary condition simplifies to Eq.\ref{eq:CircuitBoundaryCondition}. In the case of a floating boundary condition, where the impedance approaches infinity, the second term in Eq.\ref{eq:SurfaceUpdate} is ignored and the surface charge is modified through the plasma convection current $J_{conv}$ only.
An ideal voltage source reduces to a specified potential difference between the two electrodes

\begin{equation}
\phi_a - \phi_c = V(t)\textnormal{.}
\end{equation}

The last open circuit boundary model is a current driven external circuit, where assuming an ideal current source, the surface charge is driven by applying the time-varying current $I(t)$ (see Eq.\ref{eq:SurfaceUpdate}), other circuit elements are ignored \cite{Verboncoeur90}. \\
The emission of electrons from the boundary is controlled by the EmissionModel integrated into all circuit boundary models. Models are selected by specifying the model name in the \textbf{EmissionModels} list, contained within the \textbf{[case]/constant/picProperties} boundary model dictionary. Multiple models can operate simultaneously. Implemented are thermionic emission for a given temperature according to Richardson's law\cite{Crowell65} and field electron emission based on the work of Fowler and Nordheim\cite{Fowler28}.\\
A third model performs emission through sputtering of electrons by impact of heavy species, implemented similar to the BoundaryEvent sputtering model. In both implementations the user specifies the probability for the emission of selected species by impact of other species and the energy of the newly created species.\\
All emission models inject uniformly from the surface of the selected boundary patch, the models maintain this uniform distribution on separated surfaces that may occur in parallel simulations. For emission caused by species impact using the sputtering model, which occurs during the movement of the particles (see section \ref{sec:ParticlePusher}), the model automatically corrects the velocity of the particle and moves the new particle the remaining fraction of the time step. Velocities for injected particles are sampled from inverse Maxwell-Boltzmann distribution \cite{BirdsallLangdon05}.

\subsection{Collisions}
\label{sec:collision}

Plasmas which are simulated with the classic PIC scheme are collisionless, this means that forces acting on the individual particles are only a source of their macroscopic fields. As a consequence, fields generated by individual particles are excluded, meaning the field decreases with decreasing distance, and inter-particle forces inside a mesh cell are underestimated. To compensate this error picFoam implements a Coulomb collision submodule as discussed in section \ref{sec:coulombcollision}. In addition to this, picFoam also performs binary collisions between neutral species in order to account for inter-molecular forces, as well as collisions between heavy ion and neutral species, both managed in a separate submodule (see section \ref{sec:binarycollision}). A third collision submodule handles the collisions between electrons and neutral species, this submodule includes a variety of collision events and corresponding cross section models (see section \ref{sec:ElectronNeutralcollision}).\\
Besides the simulation with discrete neutral particles, picFoam also supports the simulation with a background gas model. This model is used for simulations where the neutral gas density is much higher than the plasma density, freeing memory by removing the need of simulating a lot of numeric particles and thereby greatly reducing the overall computational cost. In this model velocities of neutral particles, used in the collision process with other species, are sampled based on the Maxwell-Boltzmann distribution, using a background density and a temperature field. The background model is fully incorporated into the collision algorithms, meaning that in addition to the background models, neutral particles can be simulated at the same time, if required. All submodules mentioned are selected at run time in the \textbf{CollisionModels} dictionary in \textbf{[case]/constant/picProperties}.\\
The calculation of the new velocities while scattering are based on the assumption of equal particle weight. Since picFoam supports arbitrary weighted particles, meaning one simulated particle includes a arbitrary number of real particles, a correction has to be performed. If this correction step is neglected in the collision between particles whose weights are different, the particle with the higher weight would be scattered too much, resulting in a artificial heating of the system.
For the correction, two models have been implemented. In the simplest model, described by Nanbu and Yonemura\cite{NanbuYonemura98}, the particle with the higher weight only undergoes a collision with a probability of the ratio of the lower to the higher weight. Since this method does not conserve energy and momentum in every collision, it only works well in the case of high particle numbers. The second model implemented, described by Sentoku and Kemp\cite{SentokuKemp08}, does not have this limitation, here the heavier particle's velocity is reduced in way that energy and momentum are conserved.

\subsubsection{Coulomb collisions}
\label{sec:coulombcollision}
The Coulomb collision algorithms implemented in picFoam are based on the binary collision model introduced by Takizuka and Abe \cite{TakizukaAbe77}. In this model charged particles in a cell are paired randomly with another charged particle for collision. The idea behind this is that the long-ranged Coulomb collisions between two charged particles are negligible on distances higher than the Debye length. Which can be done because in the PIC method cell sizes are typically in the order of this length.\\
The original implementation of Takizuka's and Abe's model is only valid for charged species whose weight is equal among all particles of one species, for this reason picFoam also implements the model of Nanbu and Yonemura\cite{NanbuYonemura98} for pairing arbitrary weighted particles. Due to these models purpose of pairing charged particles, they are called \textbf{PairingAlgorithm} within in the context of the solver, which can be selected in the \textbf{[case]/constant/picProperties} dictionary. This \textbf{PairingAlgorithm} represents one part of the Coulomb collision algorithm, the second model specifies the calculation of the particle scattering process. Implemented is a model based on the work of Nanbu\cite{NanbuCoulomb97}. In it the collisions are handled in a cumulative way; a succession of small-angle collisions is grouped into a unique large scattering angle collision, making it possible to treat the collision like one between neutral atoms. Fully relativistic collisions are also implemented based on the work of Pérez et al.\cite{PerezCoulomb12}. Moreover, the solver implements the collisional ionization model for ion species described by Pérez et al.\cite{PerezCoulomb12}, which can be used in combination with both Coulomb collision models mentioned here. This model calculates an ionization cross section based on the atomic orbital binding energy, supplied by the user in the \textbf{picProperties} dictionary. Values for this can be found in references \cite{NistData, ChembioData}. An important measure in the Coulomb collision is the dimensionless Coulomb logarithm. It can be seen as the ratio of the probabilities for large-angle scattering through successive small-angle collisions over a large-angle scattering through a single Coulomb collision. It is an important parameter in the calculation of the deflection angle. By default the value of the Coulomb logarithm is calculated during collision, however, with picFoam the value can also be set to any user defined value for all different collision partners.

\subsubsection{Binary collisions}
\label{sec:binarycollision}

Binary collisions between neutral species and neutral-ion species are implemented based on the Null-Collision method developed for simulation of rarefied gases using the Direct Simulation Monte Carlo(DSMC) method described by Bird\cite{Bird94}. The collisional interaction between these species is short ranged and the scattering is treated as isotropic. In our implementation, collision probabilities are determined from a total cross section model using the hard-sphere model as approximation. 
For some simulations like DC discharges, where ions with high kinetic energy in the plasma sheath can be converted into ions with thermal energy in a charge exchange collision\cite{Nanbu00}, we implemented the simple charge exchange model for inert gases described by Nanbu and Wakayama\cite{Nanbu99}. In this model the cross section of the event is expressed by half the total cross section calculated with the hard-sphere model using a slightly higher, experimental validated, particle diameter.\\

\subsubsection{Electron neutral collisions}
\label{sec:ElectronNeutralcollision}

The last collision submodule handles collisions between electron and neutral species. Implemented are elastic, inelastic excitation, and ionization collisions, all based on one collision algorithm. In this model scattering is dependent only on electron energy, for low energies scattering is isotropic, while with increasing energies scattering takes a form of forward scattering. In inelastic collision events, precollision velocities are modified based on the threshold energy of the specific collision event, making it possible to use only one algorithm. The validity of this method and implementation details are discussed in Nanbu's work\cite{Nanbu00}.\\
The selection of the collision event is done with the help of the Null-Collision method, where the collision probability is calculated using the maximal possible collision cross section $\sigma_{max}$. In this method the collision is treated as real with a probability equal to the ratio of the probability of the single event

\begin{equation}
P_c = n_g \left|\mathbf{u}\right| \sigma \Delta t\textnormal{,}
\label{eq:CollisionProp}
\end{equation}

\noindent to the maximum probability $P_c/(P_c)_{max}$ or as not occurring with a probability of $1-P_c/(P_c)_{max}$. In Eq.\ref{eq:CollisionProp}, $n_g$ is the number density of the gas the electron collides with, $\left|\mathbf{u}\right|$ is its velocity and $\sigma$ is the cross section for the specific event. In the calculation of $(P_c)_{max}$ the variable $\sigma$ is the maximal possible cross section $\sigma_{max}$. Several cross section models for the different collision events are implemented\cite{Straub95, Wetzel87, Raju04, Brusa96}, which can be independently selected in the sub dictionary \textbf{CrossSectionModels} in the settings dictionary of the electron neutral collision model.

\subsection{Diagnostics}
\label{sec:Diagnistics}

The diagnostics submodule provides a number of models printing information on species to the standard output at run time. These models include, among others, diagnostics information on temperature, kinetic and potential field energy, momentum, and the composition of the particle cloud. Information are printed as global averages and for separate species. The selection of independent models takes place in the sub dictionary \textbf{Diagnostics} in \textbf{[case]/constant/picProperties}, optionally for every model calculation time points can be specified here, reducing the computational costs.

\section{Validation}

In this section validations for the different submodules are presented.  

\subsection{Particle pusher}
\label{sec:ValidationPusher}
The first validations are simple simulations that look at the movement of a single particle in constant electric and magnetic fields separately. After that, a more elaborated test looks at the plasma oscillation of a plasma, which validates the particle pusher and Maxwell solvers cooperative work.

\subsubsection{Electrostatic acceleration}

In this test a single electron particle is accelerated through constant electric fields calculated from fixed potential differences in a 1D mesh.\\
From energy conservation Eq.\ref{eq:EnergyConv} we retrieve the final velocity after a charge has passed through the potential difference. In this relation $m=\gamma m_e$ is the relativistic mass with rest mass $m_e$ of the electron. $v$ is the magnitude of the particle's velocity, $q$ is its electric charge, and $U$ is the potential difference the particle is accelerated through.

\begin{equation}
\begin{split}
&E_{kin} = \frac{1}{2} m \left|\mathbf{u}\right|^2 = q_pU = E_{pot} \\
&\rightarrow u = \sqrt{\frac{2q_pU}{m}}
\end{split}
\label{eq:EnergyConv}
\end{equation}

\begin{figure}
	\centering
	\includegraphics[width=0.6\textwidth]{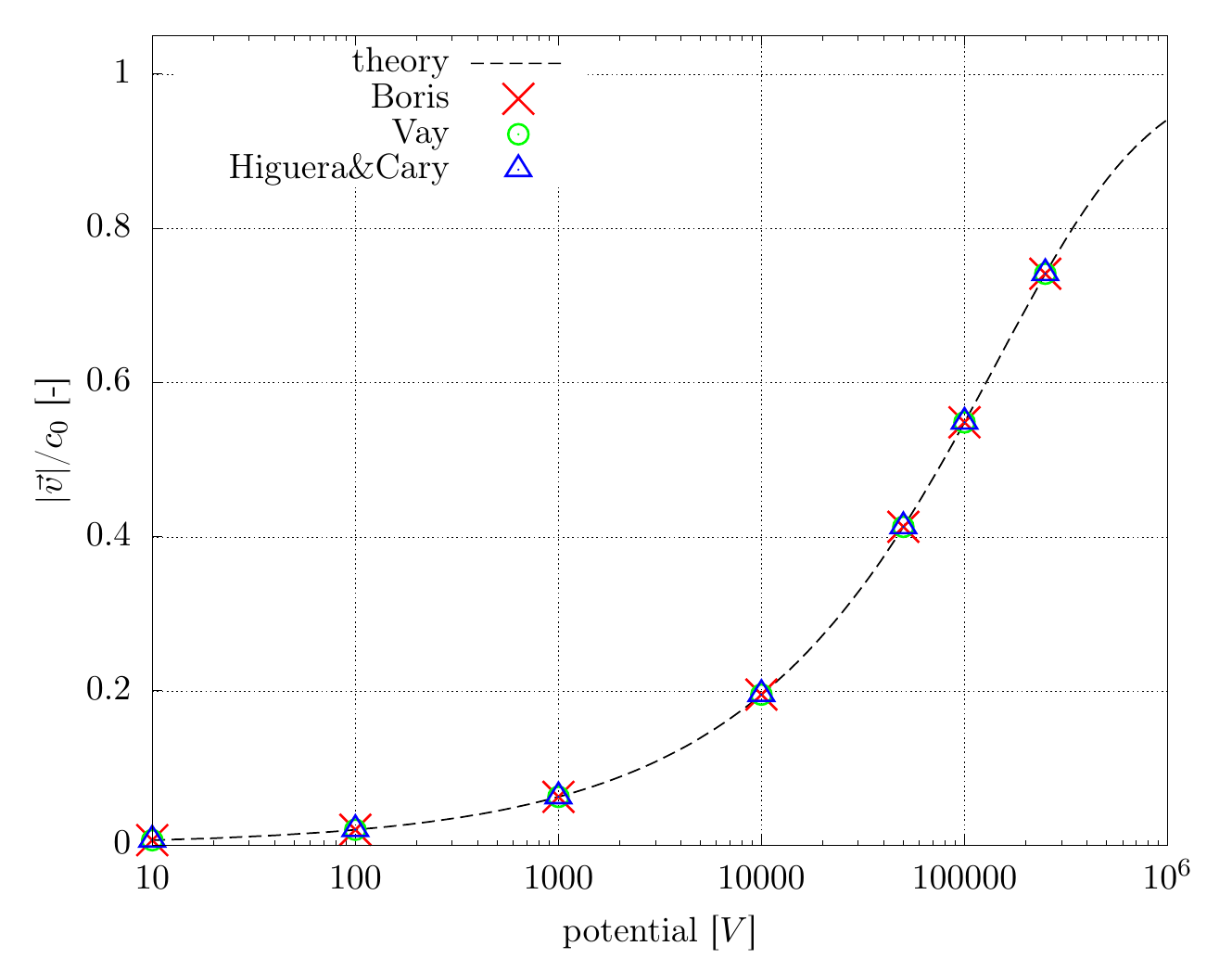}
	\caption{Final velocity of an electron accelerated through various potential differences.}
	\label{fig:ConstantAcceleration}
\end{figure}

For this test we initialize the electric field without space charge, after that an electron is created with zero velocity at the left boundary of the 1D mesh. During the simulation the Maxwell solver is turned off, since the space charge would significantly alter the electric field on simulations with low potential differences. The mesh dimensions and cell division are irrelevant, given that the Maxwell solver calculates the correct electric field strength, the particle's final velocity has to coincide with the theoretical value of Eq.\ref{eq:EnergyConv}.\\
As can be seen in Fig.\ref{fig:ConstantAcceleration} the final velocity matches the theoretical curve for all simulated potential difference.

\begin{figure}
	\centering
	\begin{subfigure}{.49\textwidth}
		\includegraphics[width=.95\textwidth]{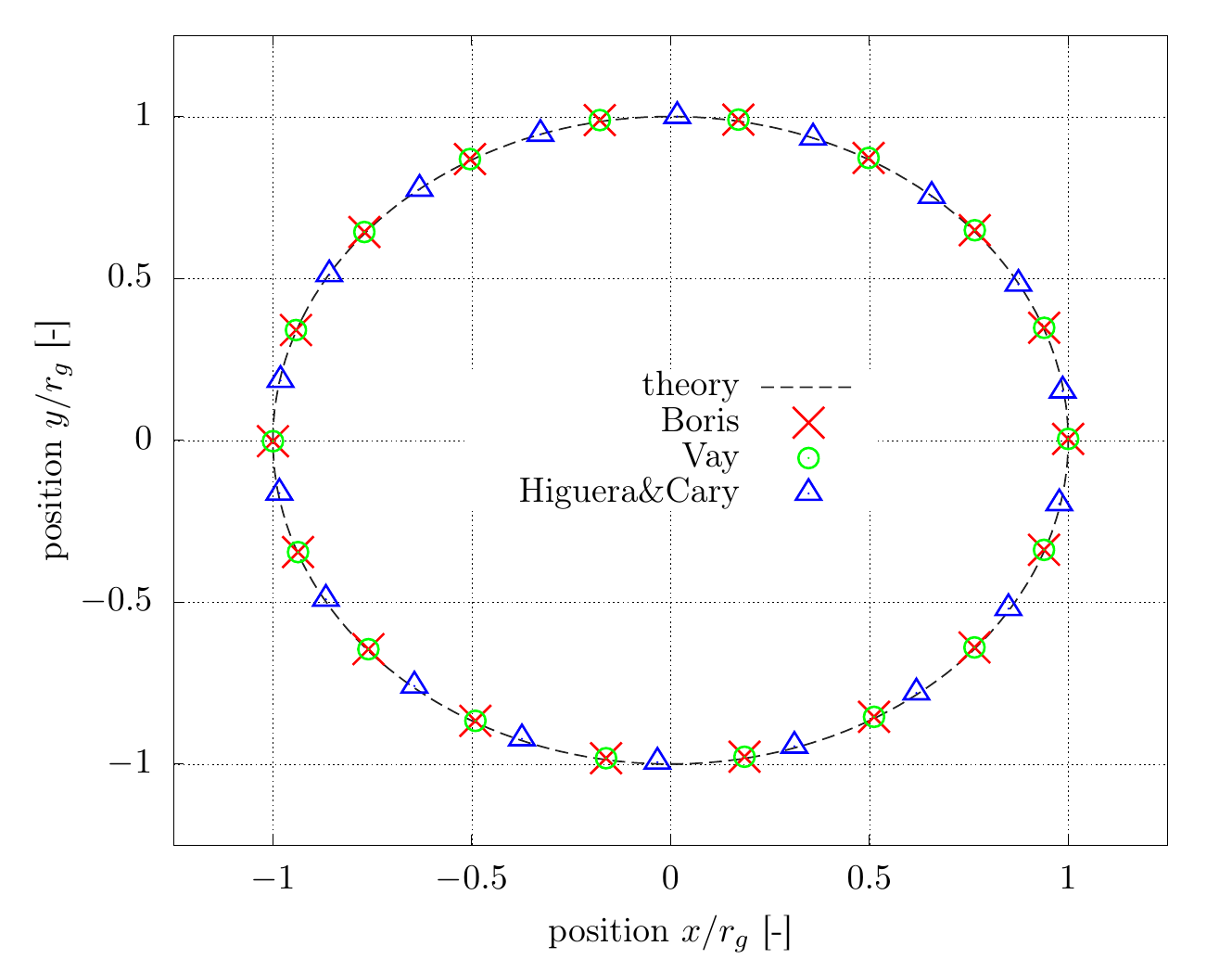}
		\caption{Trajectory of an electron after completing 1000 loops in a constant magnetic field $B_z = 1$ T.}
		\label{fig:ConstantRotation}
	\end{subfigure}
	\begin{subfigure}{.49\textwidth}
		\includegraphics[width=0.95\textwidth]{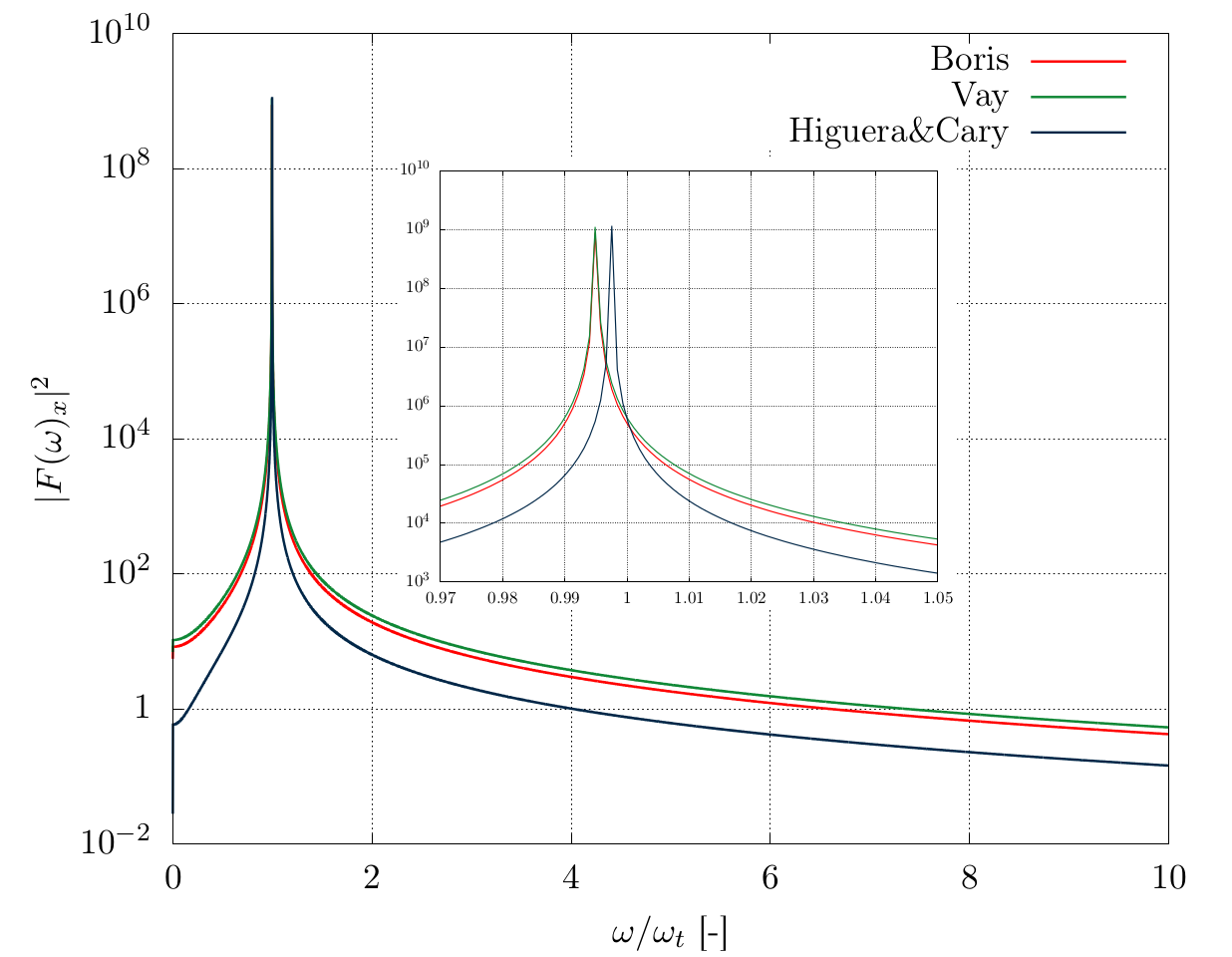}
		\caption{Power spectra of the Fourier transformation of the electron's x-coordinate in time. The inner graph shows the zoom around the tips.}
		\label{fig:RotationFFT}
	\end{subfigure}
	\caption{Results for the validation of section \ref{sec:MagneticRotation}}
\end{figure}

\subsubsection{Rotation in magnetic field}
\label{sec:MagneticRotation}
Here we validate the implementation of the magnetic rotation of all models. To do so we compare the circular trajectory of a single electron in a constant magnetic field to its theoretical circular trajectory with the radius of gyration given by $r_g = m\,u_\perp/\left|q_p\right|\,B$.
A single electron is placed at $\mathbf{x} = (0, -r_g, 0)^T$ with a velocity of $u_\perp = 0.1 c_0$ in a 2D rectangular mesh of $400 \times 400\,\textnormal{mm}^2$  with 40 cells in every valid dimension. The Maxwell solver is turned off during the simulation, the time step is set to $0.1$ fs and the simulation ends after 1000 rotations. The magnetic field $B_z = 1$ T is constant.\\ 
Fig.\ref{fig:ConstantRotation} shows the final trajectory after 1000 loops and its agreement with theory. To explain the deviation of Higuera and Cary's model from the other two, we conducted a analysis of the angular frequency $\omega_l = \left|q_p\right|\,B/m$. For this we compute the Fourier transformation of the particle's x-coordinate in time. Fig.\ref{fig:RotationFFT} shows the power spectra of the Fourier transformation. The peak for Boris and Vay's method is at 27.85 GHz, giving an error below 1\% to the theoretical value of 28 GHz. Note that the sample frequencies are multiple of 25 MHz. For Higuera and Cary's method the Fourier analysis shows a peak at 27.925 GHz giving a smaller error than Boris and Vay's methods, this showing that all methods implemented are in very good agreement with theory.

\subsection{Plasma oscillation}
\label{sec:PlasmaOscillation}
In this validation we look at the plasma oscillation of electron species deflected from a uniform ion background, validating the particle movement in combination with the computation of the electric field. For this test 1.536 million ion parcels with a weight of $\omega_p = 2.5\cdot 10^6$ are uniformly distributed in $6\,\textnormal{m}\times0.8\,\textnormal{m}\times0.8\,\textnormal{m}$ wide mesh with $120\times16\times16$ cells. From the location of the ion parcels the same number of electrons, deflected in the x-direction by $\Delta x = A \sin({k x_1})$ with an amplitude of $A=0.1$ m and a wavenumber of $k=4\pi/6$ m, is placed inside the mesh. This leads to a number density of $n_e = 10^{13}\,\textnormal{m}^{-3}$, according to theory\cite{Bittencourt04} the electrons oscillated with an angular frequency of $\omega_{pe} = \sqrt{n_e e^2/(\epsilon_0 m_e)}$, where $m_e$ is the rest mass of an electron, giving us a period of $T=2\pi/\omega_{pe} = 35.22$ ns. Initially, both species are at rest, during the simulation the ion species is kept fixed, while the electron species starts oscillating accelerated by the potential gradient induced by the deflection from its equilibrium position at the ions. The emerging plasma frequency is determined from two periods of the total kinetic $\sum 0.5 \gamma m_e \left|\mathbf{u}\right|^2$ and field energy $\sum 0.5 \epsilon_0 \left|\mathbf{E_c}\right|^2 / V_c$, due to the electron having to pass the equilibrium position twice for a whole oscillation period \cite{Stock13}. Here $V_c$ is the cell volume and $\mathbf{E}_c$ is the electric field vector defined at the cell center.

\begin{figure}[h]
	\centering
	\includegraphics[width=0.6\textwidth]{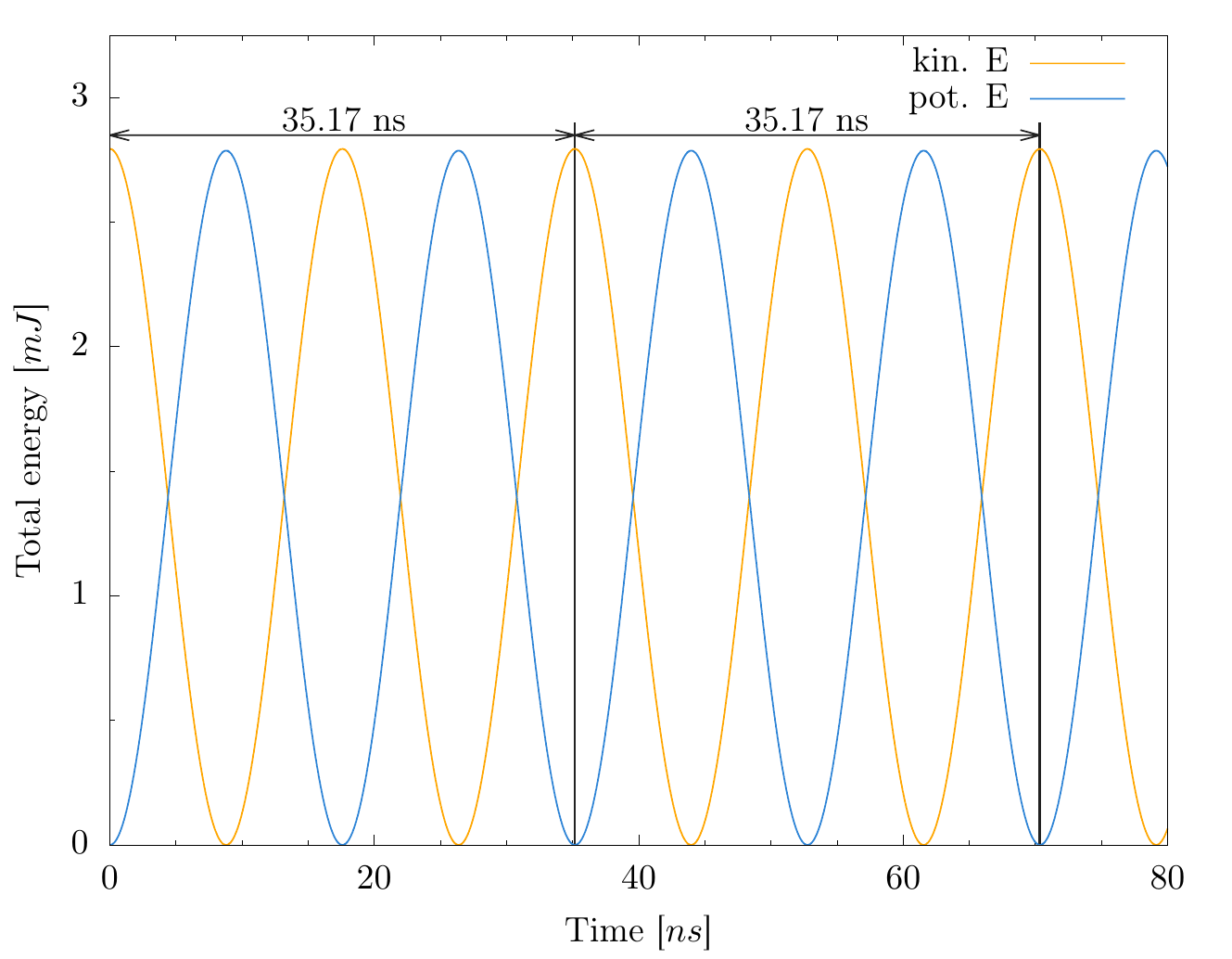}.
	\caption{Plot of the particles total kinetic $\sum 0.5 \gamma m_e |\mathbf{u}|^2$ and the corresponding electric fields total potential energy $\sum 0.5 \epsilon_0 |\mathbf{E_c}|^2/ V_c$ over time. Two cycles through the equilibrium position correspond to the plasma frequency $\omega_{pe}$, with its period $T=2\,\pi/\omega_{pe}$. }
	\label{fig:oscillationPlot}
\end{figure}

Fig.\ref{fig:oscillationPlot} shows the energy plot over time. The evaluation of the results is presented in Tbl.\ref{tbl:oscillationResult}, showing an error for the oscillation period below 1\% for both implemented weighting methods. The second and third row of the table show an evaluation of the energy conservation, here the error represents the average deviation from the first peak of potential and kinetic energy, revealing that the volume weighting method (section \ref{sec:VolumeWeighting}) provides conservation of energy two orders of magnitude higher than the cell average method (section \ref{sec:CellAverageWeighting}).
The last row shows the evaluation of momentum conservation, here we look at the trend of the total momentum of the system in x-direction. For the simulation parameter mentioned above, leading to an average number of N$^C_P=50$ particles per cell per species and using the cell average method we observe an unstable behavior. Which is further increasing over time due to the instabilities induced by noise in the charge density field. For the volume weighting method a stable oscillation in the total momentum is observed with an error of 2.8\% per period of the plasma frequency. Increasing N$^C_P$ to 100 reduces the error for the volume weighting method to 1.61\%, while the behavior using the cell average method stays unstable due to prevailing noise.\\

\begin{table}[h]
	\centering
	\begin{tabular}{l||r|r}
		& CellAverage & VolumeWeighting  \\ \hline \hline
		period T[ns] & $35.01$  & $35.17$ \\
		error [\%] & $0.59$  & $0.14$ \\ \hline
		max $E_{pot}$ [mJ]&  2.797 & 2.795 \\ 
		error $\Delta E_{pot}$ [\%] & 0.96 & 0.0065\\ \hline
		max $E_{kin}$ [mJ]& 2.797 & 2.788 \\ 
		error $\Delta E_{kin}$ [\%] & 0.53 & 0.0057\\ \hline
		error $\Delta p_{x}$ [\%] (N$^C_p$ = 50)& -- & 2.82\\
		error $\Delta p_{x}$ [\%] (N$^C_p$ = 100)& -- & 1.61\\
	\end{tabular}
	\caption{Computed period of the plasma oscillation using different weighting schemes and the error to the theoretical value of $T=2\pi/\omega_{pe} = 35.22$ ns. The second and third row show the energy conservation by the average deviation from the first peak of the total kinetic and potential field energy at N$^C_P = 50$. The last row shows the error per oscillation period in the momentum conservation for different number of particles per cell. }
	\label{tbl:oscillationResult}
\end{table}

\begin{figure}
	\centering
	\begin{subfigure}{.49\textwidth}
		\includegraphics[width=.99\textwidth]{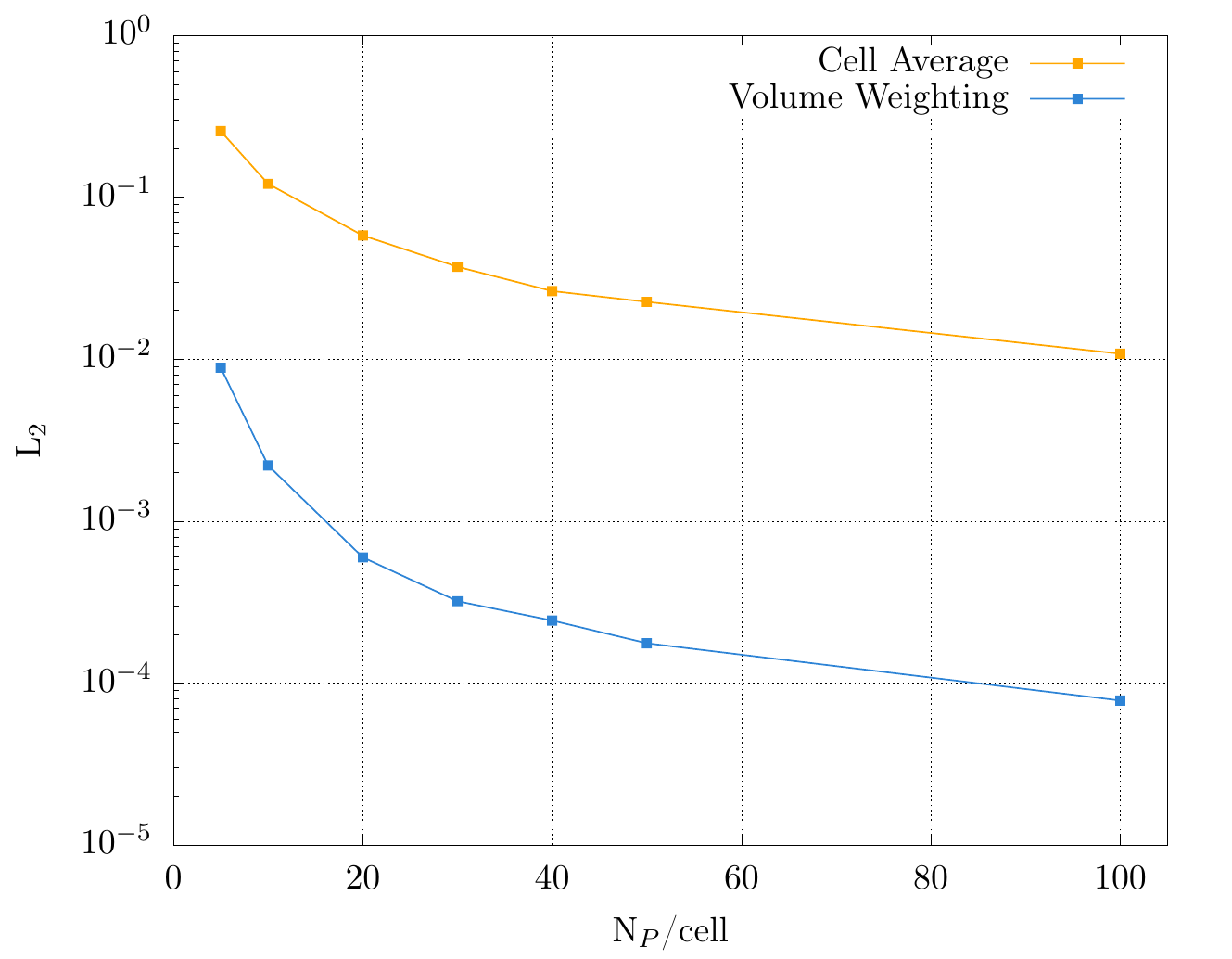}
		\caption{L$_2$ norm of the relative error in the charge density field plotted over the number of particles N$_P$ per cell.}
		\label{fig:NoiseOverNp}
	\end{subfigure}
	\begin{subfigure}{.49\textwidth}
		\includegraphics[width=0.99\textwidth]{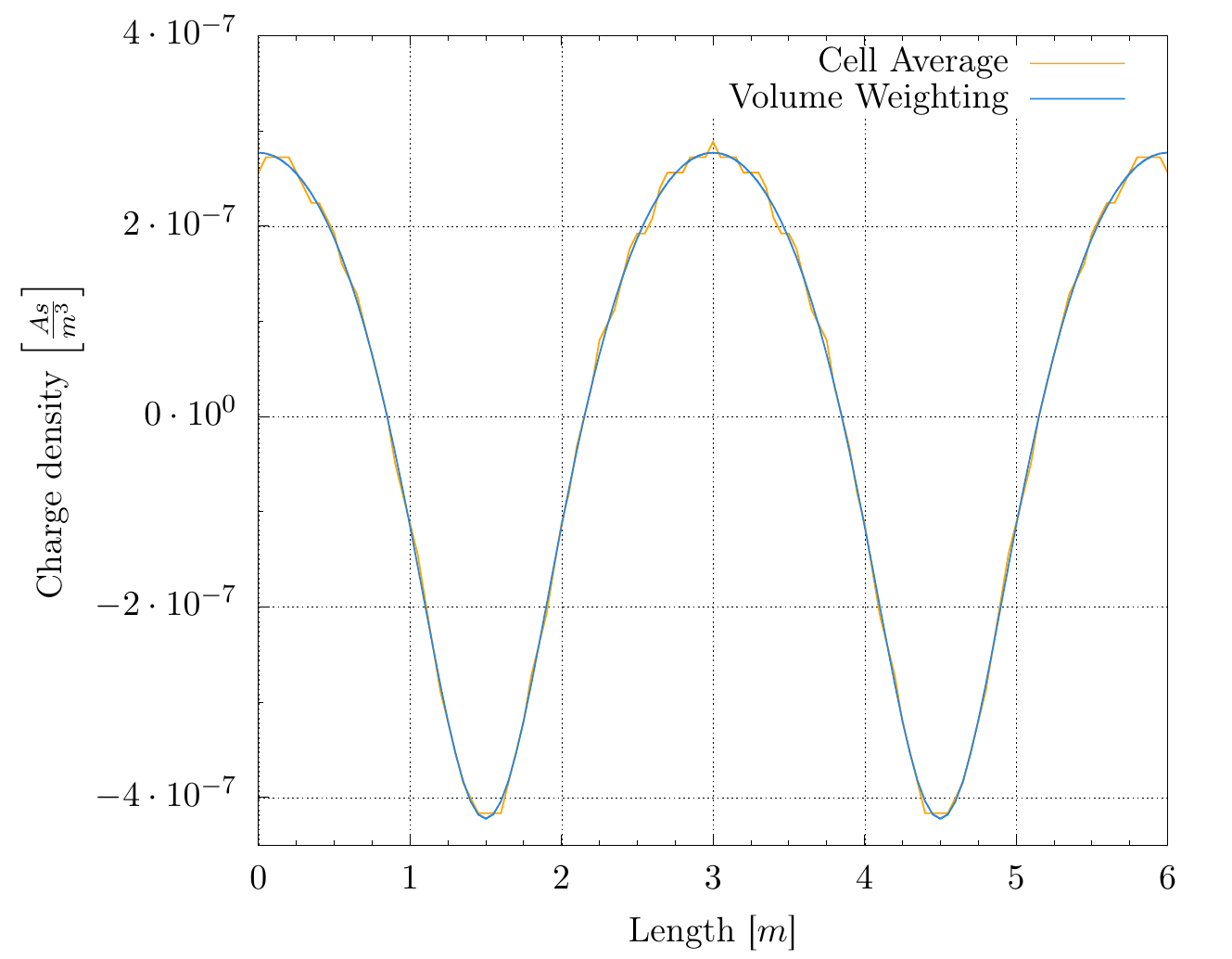}
		\caption{Initial charge density over the domain length of the simulation in section \ref{sec:PlasmaOscillation} at N$^C_P=50$, here showing noise when using the cell average method compared to the volume weighting method.}
		\label{fig:NoiseNp50}
	\end{subfigure}
	\caption{Investigation of noise in the charge density field for the implemented weighting methods.}
\end{figure}

An accurate representation of the charge density field is an important factor in the correct description of the plasma. In this, the weighting method has a big impact on the noise of the field. Fig.\ref{fig:NoiseNp50} shows the charge density field in the x-direction of the simulation with N$^C_P=50$. As can be seen, compared to the volume weighting method, the cell average method introduces slight noise to the field. To quantify the noise introduced by the different methods, we compare the initial fields calculated by both methods for different numbers of particles per cell in the range of 5 to 100 to the initial field produced by the volume weighting method with 1000 particles per cell. By calculating the L$_2$ norms of the relative errors we obtain an indication of the accuracy of the weighting method. Fig.\ref{fig:NoiseOverNp} shows that the field calculated with 5 particles per cell and the volume weighting method is more accurate than the field calculated for 100 particles per cell and the cell average method. Therefore, despite the higher numerical cost, it is recommended to use the volume weighting method for simulations conducted with picFoam. While both methods accurately conserve energy, the cell average method tends to inaccurately conserve momentum because of noise introduced to the charge density field. Over time this will lead to a loss of physical fidelity. Similar was observed by Capon et al. \cite{Capon2017}, which implemented a cell average method and a related volume weighting method and showed an inaccuracy in the velocity of a single electron oscillating around an ion, when using the cell average method.

\subsection{Collision and Ionization} 
\label{sec:ValidationCollision}
To validate the collision algorithms implemented in the new code, simulations are run in which two species are initialized with different temperatures but with no relative drift, in this case their equilibration is described by

\begin{equation}
\frac{dT_\alpha}{dt} = \sum_{\beta} \overline{\nu}^{\,\alpha|\beta} (T_\beta - T_\alpha)\textnormal{.}
\label{math:theoryTprofile}
\end{equation}

\noindent Where $\overline{\nu}^{\,\alpha|\beta}$ is the mean collision frequency of species $\alpha$ with species $\beta$ \cite{nrlPlasma19}.

\subsubsection{Neutral-Binary collisions}
\label{NeutalBinaryValidation}
The first collision submodule handles collisions between neutral species as well as collisions of neutral with ion species. The interaction is short range, we treat the collision by using the hard-sphere model. Scattering is isotropic. The mean collision frequency for species $\alpha$ with species $\beta$ is described by $\overline{\nu}^{\,\alpha|\beta} = n_\beta \overline{g} \sigma_T$, where $\overline{g}$ is the mean relative velocity, $n_\beta$ the number density of species $\beta$, and $\sigma_T$ the hard-sphere cross section $\pi (d_\alpha+d_\beta)^2/4$, which is constant. For the validation of this submodule argon and neon atoms with an atomic mass of $39.948$ u and $20.18$ u, and van der Waals radii of 188 and 154 pm are used. Particles are equably initialized with a number density of $10^{22}\,\textnormal{m}^{-3}$ respectively on a 15$\,\mu\textnormal{m}\times$15$\,\mu\textnormal{m}$ 2D mesh with a total of 64 cells. Velocities are sampled from a Maxwell-Boltzmann distribution with argon begin at 1000 K and neon at 500 K. Three simulations with weighting ratios ($\Delta \omega_p = \omega_\alpha /\omega_\beta$) of 1, 2, and 5 were performed, here the weight of $\omega_\alpha$ is kept constant at 400. The correction method for differently weighted particles described by Nanbu and Yonemura is used \cite{NanbuYonemura98}.The theoretical collision frequency is calculated with the mean relative velocity of $\overline{g} = \sqrt{8 \overline{T} k_B/(\mu \pi) }$, where $\overline{T}= T_\infty = 750$ K, and $\mu$ is the reduced mass of both species.

\begin{figure}
	\centering
	\includegraphics[width=0.6\textwidth]{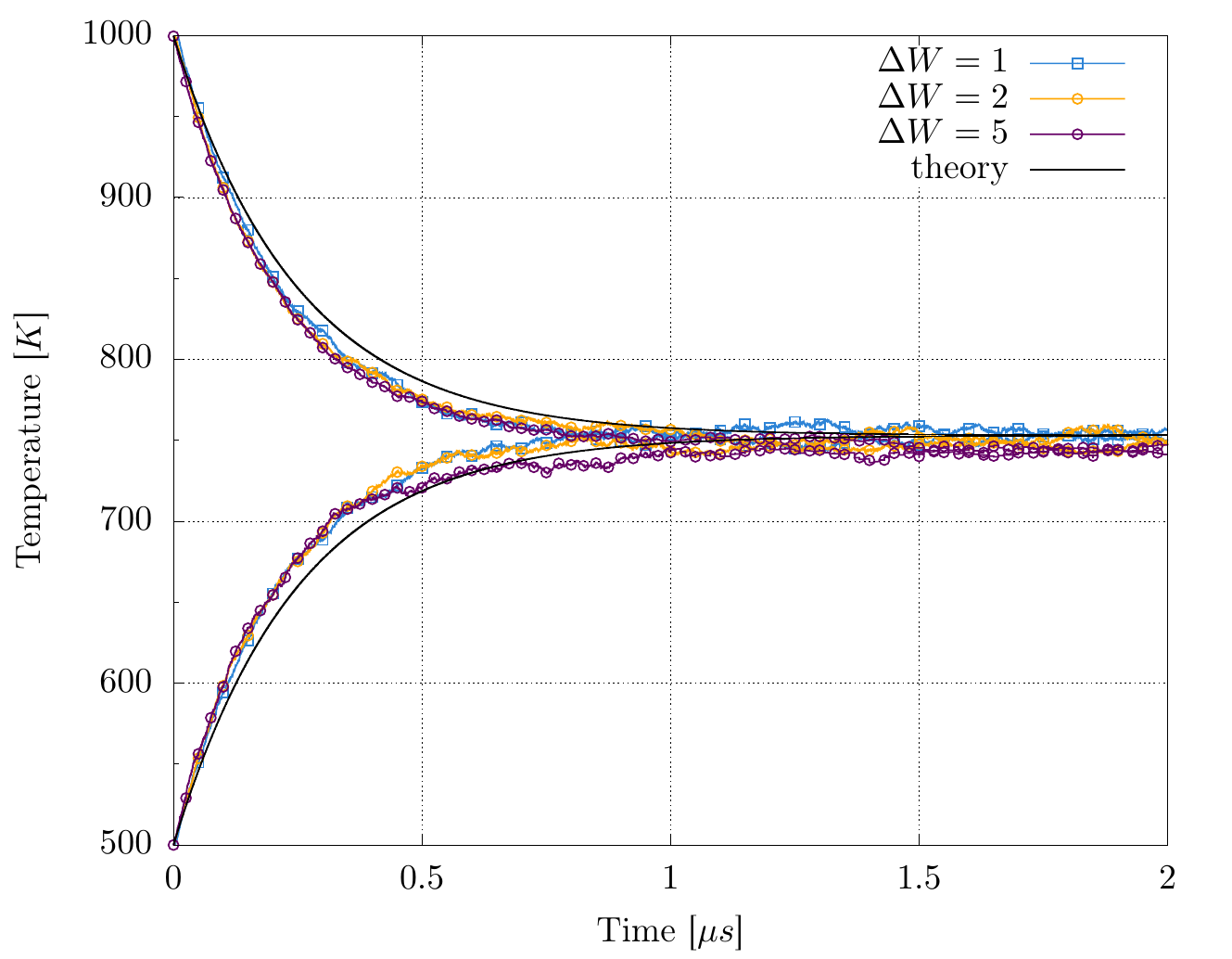}.
	\caption{Thermalisation of species argon and neon trough binary collisions, simulated with three particle weight differences using the correction scheme of Nanbu and Yonemura\cite{NanbuYonemura98}.}
	\label{fig:binaryCollisionPlot}
\end{figure}

\noindent Fig.\ref{fig:binaryCollisionPlot} shows the temperature profile, reasonable agreement with the theory was achieved. Small deviations can be explained by the use of an mean relative velocity $\overline{g}$. 

\subsubsection{Coulomb collisions}

This test uses the model described by Perez et al.\cite{PerezCoulomb12} and is set up with the same mesh as the validation in section \ref{NeutalBinaryValidation}. 160000 real electrons and argon ions are distributed in a way that each cell has the same number of particles so that the plasma is quasineutral with number densities of $n_e=n_i\approx3.034\cdot 10^{22}\,\textnormal{m}^{-3}$. Weight ratios are set to 1, 2, and 5 as before while keeping the weight of the ions constant at 400. The weighting correction method used here is the model described in Sentoku and Kemp\cite{SentokuKemp08}. For faster convergence the mass ratio of the species $\Delta m = m_i/m_e$ is set to 10, where $m_e$ remains constant at the rest mass of electron $m_e = 9.109 \cdot 10^{-31}$ kg.

\begin{equation}
\overline{\nu}^{\,\alpha|\beta} = 1.8 \times 10^{-19} \frac{Z_\alpha^2 Z_\beta^2 \,n_i \sqrt{m_e m_i} \ln\Lambda}{(m_e T_e + m_i T_i)^{3/2}}
\label{math:coulombColFreq}
\end{equation}

\begin{figure}
	\centering
	\includegraphics[width=0.6\textwidth]{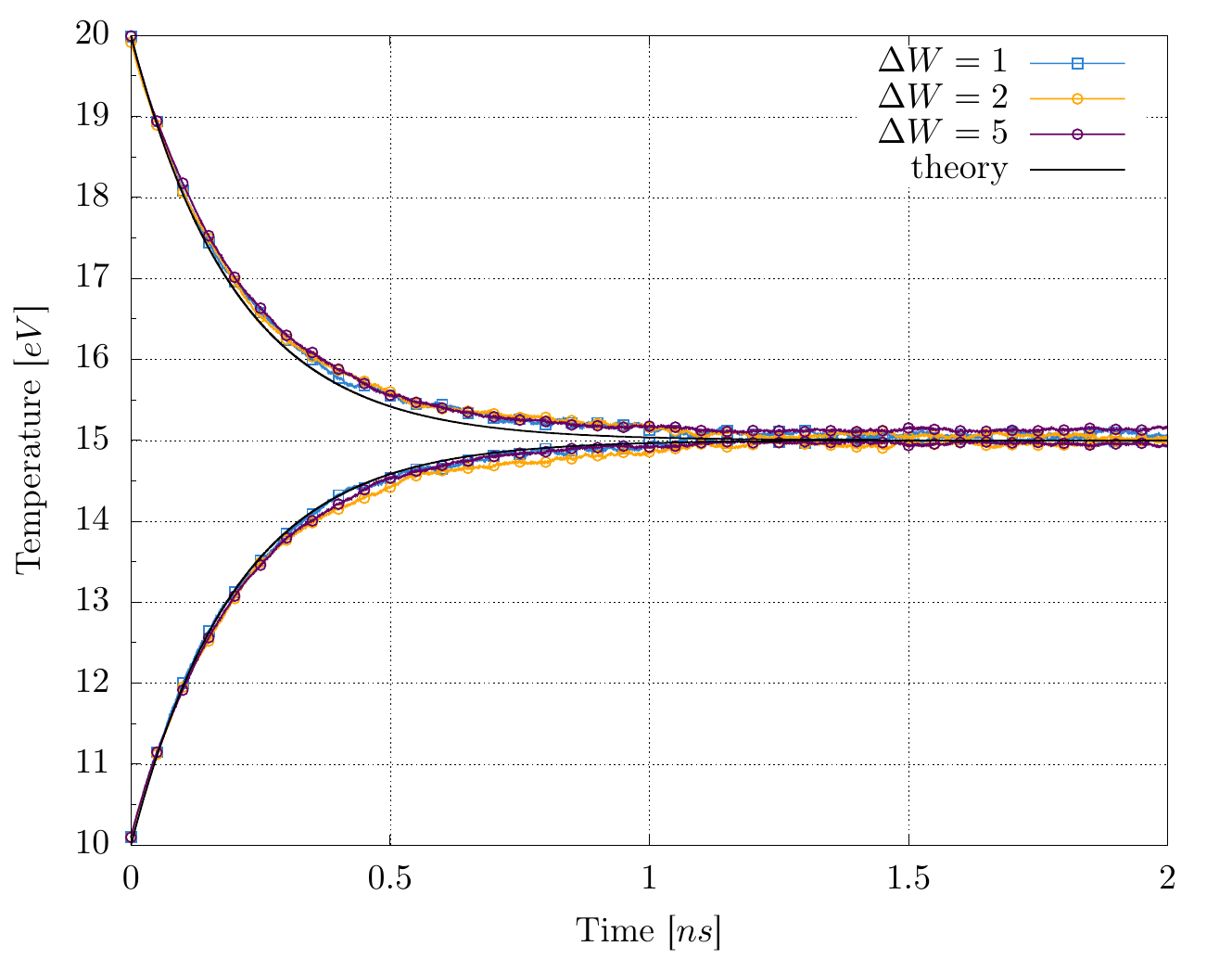}.
	\caption{Thermalisation of electrons and argon ions trough relativistic Coulomb collisions\cite{PerezCoulomb12}, simulated with three particle weight differences using the correction scheme of Sentoku and Kemp\cite{SentokuKemp08}.}
	\label{fig:coulombCollisionPlot}
\end{figure}

\noindent As in the last section the theory is described by Eq.\ref{math:theoryTprofile}, this time using the collision frequency stated by the NRL Plasma Formulary\cite{nrlPlasma19} shown in Eq.\ref{math:coulombColFreq}. Initial temperatures are set to $T_e = 20$ eV for the electrons and $T_i = 10$ eV for the argon ions, the Coulomb logarithm is fixed to $\ln \Lambda = 15$ for all possible collision partners. Fig.\ref{fig:coulombCollisionPlot} shows the temperature profiles over time and an excellent agreement with the theory for all simulations.

\subsubsection{Ionization}

The last test for validation of the collision submodules looks at the ionization rate of a neutral argon species during electron impact. This allows for a validation independent of the other submodules as a thermalization simulation requires the other submodules for inter species collision.\\
The same mesh and the three weight ratios as in the previous tests are used. Both species are initialized with a number density of $10^{22}\,\textnormal{m}^{-3}$, the weight of the argon species is fixed at 100. Ionization is handled with a model described by Nanbu\cite{Nanbu00}. Since the ionization cross section is dependent on the electron's kinetic energy and this test aims to validate the algorithm, a fixed cross section of $3.5e^{-20}\,\textnormal{m}^2$ is used. This allows us to use Eq.\ref{math:theoryTprofile} for the description of the ionization, where instead of the temperature the number of neutral and ion particles created are used. The ionization frequency is  $\overline{\nu}^{\,\alpha|\beta} = n_\beta \overline{g} \sigma_i$ with $\overline{g} = \sqrt{8 T_e k_B/(m_e \pi)}$ the mean velocity calculated at the initial electron temperature of 100 eV.

\begin{figure}
	\centering
	\includegraphics[width=0.6\textwidth]{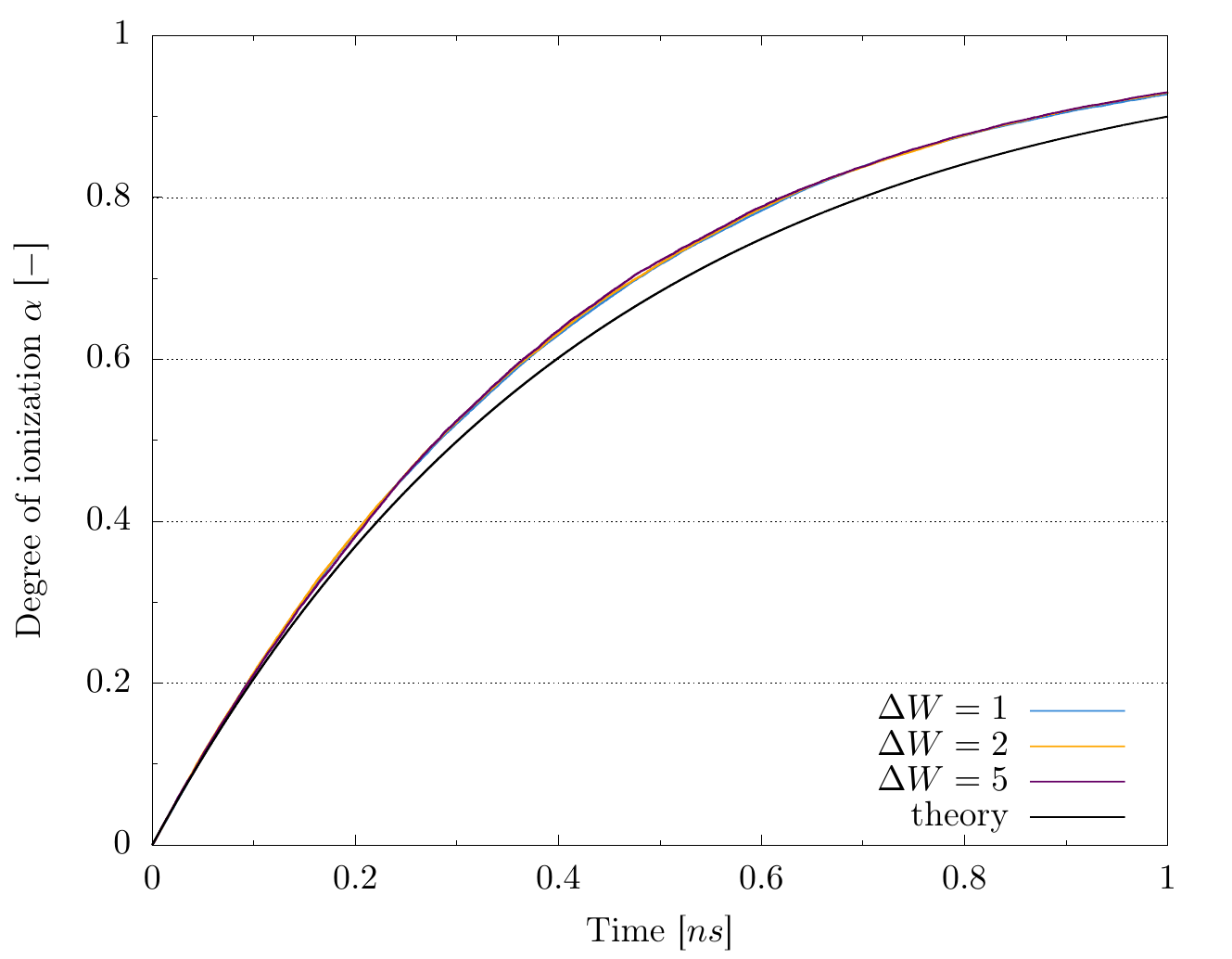}.
	\caption{Ionization of argon species in collision with electrons at a number density of $10^{22}\,\textnormal{m}^{-3}$ using a fixed cross section $3.5e^{-20}\,\textnormal{m}^2$ corresponding to the initial electron energy of $100$ eV.}
	\label{fig:ionizationPlot}
\end{figure}

\noindent Good agreement with theory can be observed in Fig.\ref{fig:ionizationPlot}. The deviation from the theoretical curve seen in all simulations here can be accounted to a changing velocity distribution at later times.

\subsection{Boundary model}
\label{sec:BoundaryValidation}
\subsubsection{Open circuit models}

\begin{figure}[h]
	\centering
	\includegraphics[width=0.6\textwidth]{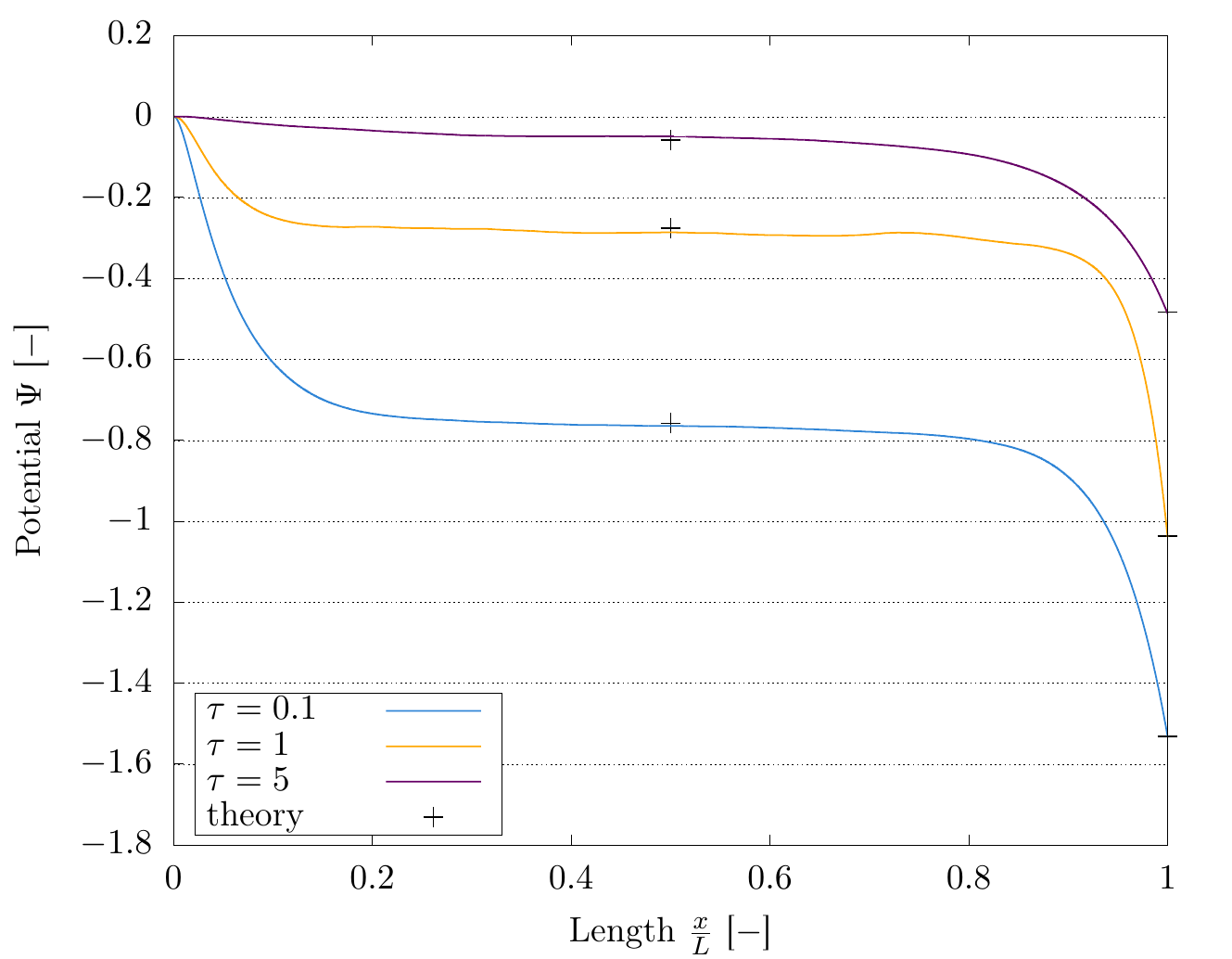}
	\caption{Plot of the potential between the plasma source and the collector for all simulated cases ($\tau=0.1,1,5$). The crosses show the theoretical values calculated from Eq.\ref{math:schwagerEq1} and Eq.\ref{math:schwagerEq2}.}
	\label{fig:potentialPlots}
\end{figure}

In this validation we look at the potential drop between a plasma and a collector represented by the floating boundary condition and compare it against theory. The open circuit boundary models, including the floating potential model, are implemented based on the gradient boundary condition Eq.\ref{eq:CircuitBoundaryCondition} and Eq.\ref{eq:SurfaceUpdate}, hence this test validates all implementations of open circuit models.\\
Schwager et al.\cite{Schwager90} derived two equations which together describe the source sheat potential drop $\psi_p$  and collector potential $\psi_c$ in terms of mass $\mu = m_e/m_i$ and temperature $\tau = T_{si}/T_{se}$ ratios of the species, here $\psi = e\phi/T_{se}$ is the, with the electron source temperature $T_{se}$, normalized potential. The first equation is derived at the inflection point in the middle of the characteristic curve, as can be seen in Fig.\ref{fig:potentialPlots}, at this point $\Delta \psi_p = 0$.

\begin{equation}
\frac{1}{\sqrt{\mu \tau}}\, \textnormal{exp} \Big(\frac{-\psi_p}{\tau}\Big)\textnormal{erfc}\Big(\frac{-\psi_p}{\tau}\Big)^{1/2} = \textnormal{exp}(\psi_p-\psi_c)[1+\textnormal{erf}(\psi_p-\psi_c)^{1/2}]
\label{math:schwagerEq1}
\end{equation}

\noindent The second equation results from imposing the zero electric field condition at the inflection point, written in separate terms of normalized integrated densities $\zeta$ for electrons and ions.

\begin{equation}
\zeta_i + \zeta_e = 0
\label{math:schwagerEq2}
\end{equation}

\noindent with

\begin{equation*}
\zeta_i = \sqrt{\frac{\tau}{\mu}}\Big[\textnormal{exp}\Big(\frac{-\psi_p}{\tau}\Big)\textnormal{erfc}\Big(\frac{-\psi_p}{\tau}\Big)^{1/2} - 1 + \Big(\frac{-4\psi_p}{\pi\tau}\Big)^{1/2}\Big]
\end{equation*}

\noindent and

\begin{equation*}
\begin{split}
\zeta_e &= \textnormal{exp}(\psi_p-\psi_c)[1+\textnormal{erf}(\psi_p-\psi_c)^{1/2}] \\
& +(2/\sqrt{\pi})(-\psi_c)^{1/2}-(2/\sqrt{\pi})(\psi_p-\psi_c)^{1/2} \\
& -\textnormal{exp}(-\psi_c)[1+\textnormal{erf}(-\psi_c)^{1/2}]\textnormal{.}
\end{split}
\end{equation*}

\noindent The simultaneous solutions of Eq.\ref{math:schwagerEq1} and Eq.\ref{math:schwagerEq2} exist in two points, one occurs at $d\psi_c /d \psi_p = 0$, the other at $\psi_p=0$. Here, the first solution is stable and represents the theoretical solution to the problem we compare to. For more details see Schwager et al.\cite{Schwager90}.\\
Tbl.\ref{tbl:potentialdrops} compares the theoretical solutions of $\mu = 1/40$ and $\tau =$ 0.1, 1, and 5, to the solution of a one-dimensional simulation using the same parameter. Initially particles are distributed with a number density of $n = 10^{18}\,\textnormal{m}^{-3}$ in the one-dimensional domain, which has a length of $l = 132\,\lambda_D$ for $\tau = 0.1$, $l = 44\,\lambda_D$ for  $\tau = 1$, and $66\,\lambda_D$ for $\tau = 5$, where $\lambda_D$ is the corresponding Debye length. During simulation particles are injected from the source with a half Maxwell-Boltzmann distribution and a rate of $5.3 \cdot 10^{12}$ Hz. Particles reflected back to the source are re-injected with the source temperature. In Tbl.\ref{tbl:potentialdrops} the collector value has been read directly from the averaged field value at the boundary, while the value for $\psi_p$ has been averaged over the flat part of the curve (see Fig.\ref{fig:potentialPlots}).
As can be seen, the error for the collector potential is below 1\% for all cases, only the $\psi_p$ for $\tau = 5$ shows a considerable error of 10\%.

\begin{table}[h]
	\centering
	\begin{tabular}{l||c|c|c|c}
		& \multicolumn{2}{c|}{Simulation}  & \multicolumn{2}{c}{Theory}  \\\hline 
		{$\tau$}\textbackslash{$\psi$} & $\psi_p$ & $\psi_c$	& $\psi_p$ & $\psi_c$ \\ \hline \hline
		0.1 &  -0.7596 & -1.532 & -0.758 & -1.531 \\
		1 &  -0.278 & -1.0389 & -0.276 & -1.036 \\
		5 &  -0.0525 & -0.487 & -0.0582 & -0.483 \\
	\end{tabular}
	\caption{Comparison of the simulated versus theoretical value of the potential at the inflection point $\psi_p$ and the collector $\psi_c$ at three different temperature ratios $\tau$.}
	\label{tbl:potentialdrops}
\end{table}

\subsubsection{General RLC circuit model}

For the validation of the RLC circuit boundary model, we perform two simulations. First, we compare the current progression of an RLC circuit driven by a sinusoidal voltage source over time with a theoretical solution. This validation is set up similar to the validation performed by Verboncoeur et al.\cite{Verboncoeur90}. In this simulation the plasma region is assigned to a permittivity of $10^{20}$ with no plasma and the circuit elements are set to $\textnormal{R} = 1\,\Omega$, $\textnormal{L} = 1\,\mu\textnormal{H}$ and $\textnormal{C} = 5\,\mu\textnormal{F}$. The voltage source follows the expression $V(t)=V_0\sin(\omega t + \theta)$, where $V_0$ is set to $1$ V, the angular frequency of $10^6$ is used and the initial phase is set to zero. The time step is set to $\Delta t = 2\pi/128\omega$. As stated in Verboncoeur et al.\cite{Verboncoeur90} the theoretical solution can be predicted using

\begin{equation}
\label{eq:RLCtheoryCurrent}
\begin{split}
&I(t)= \frac{a_2 V_0 /(\omega Z) \cos (\theta-\delta)+V_0/Z\sin(\theta-\delta)}{a_2/a_1-1}\exp(a_1 t)\\&+ \frac{a_1 V_0 / (\omega Z) \cos(\theta-\delta)+V_0/Z\sin(\theta-\delta)}{a_1/a_2-1}\exp(a_2 t)\\&+ \frac{V_0}{Z}\sin(\omega t + \theta - \delta)\textnormal{,}
\end{split}
\end{equation}

\noindent where

\begin{equation*}
\begin{split}
&Z = \sqrt{R^2 + \Big(\omega L-\frac{1}{\omega C }\Big)^2}\textnormal{,}\\
& \delta = \textnormal{asin}\Big(\frac{\omega L - 1/(\omega C)}{Z}\Big)\textnormal{, and}\\
& a_{1,2} = -\frac{R}{2L} \pm\sqrt{\frac{R^2}{4 L^2}-\frac{1}{LC}}\textnormal{.}
\end{split}
\end{equation*}

\begin{figure}[h!]
	\centering
	\includegraphics[width=0.6\textwidth]{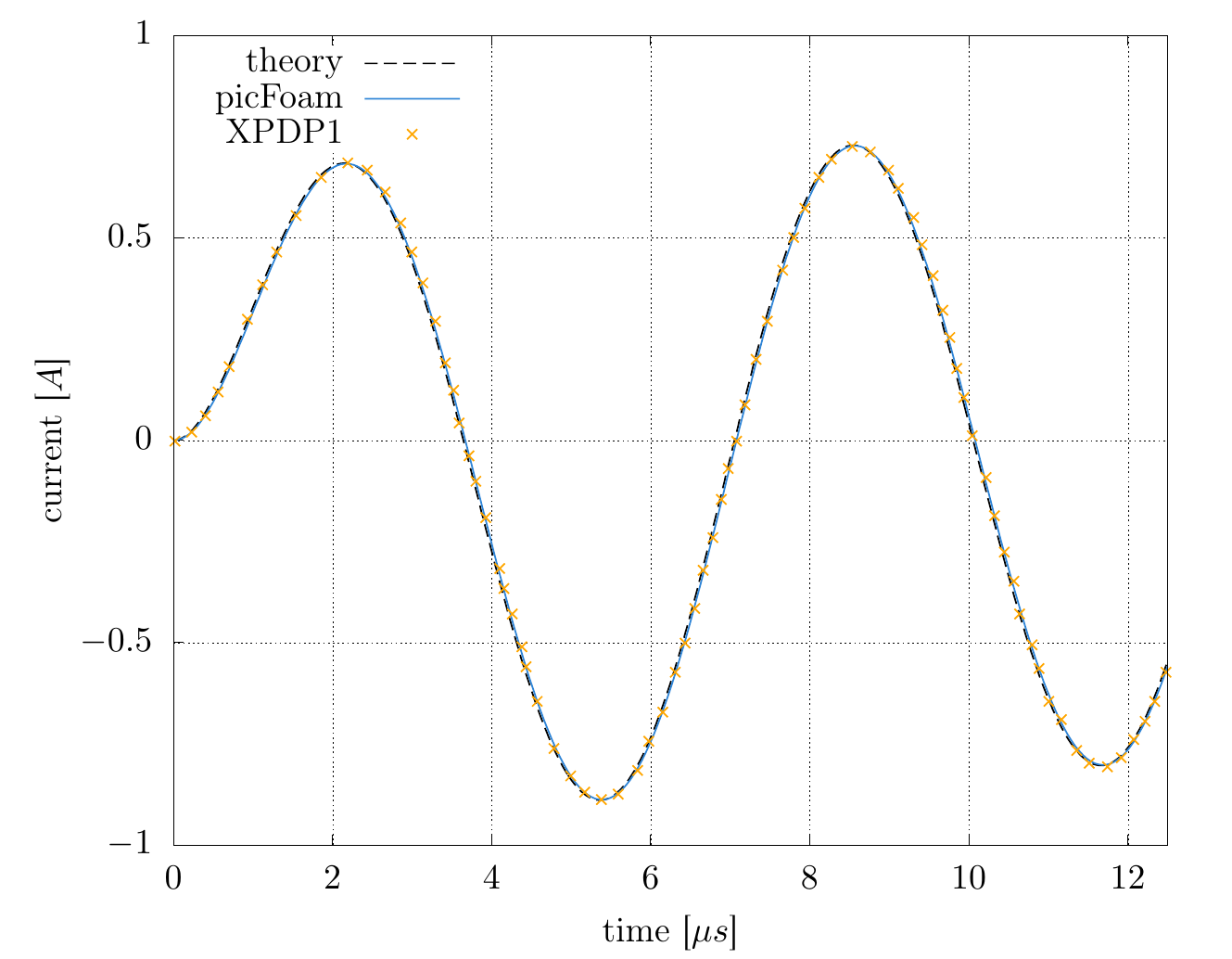}.
	\caption{Comparison of the theoretical progression of current Eq.\ref{eq:RLCtheoryCurrent} in an RLC circuit with $\textnormal{R} = 1\,\Omega$, $\textnormal{L} = 1\,\mu\textnormal{H}$ and $\textnormal{C} = 5\,\mu\textnormal{F}$ to the boundary model implementations in picFoam and PDP1 with no plasma and a permittivity of $\epsilon=10^{20}$.}
	\label{fig:RLCtheory}
\end{figure}

Fig.\ref{fig:RLCtheory} shows the current plot over time up to $\textnormal{t}=256\Delta t$. As shown here, the solution calculated by picFoam coincides with the one of PDP1 in Verboncoeur et al.\cite{Verboncoeur90}, and both agree well to the theoretical solution of Eq.\ref{eq:RLCtheoryCurrent}.

\begin{figure}[h!]
	\centering
	\includegraphics[width=0.6\textwidth]{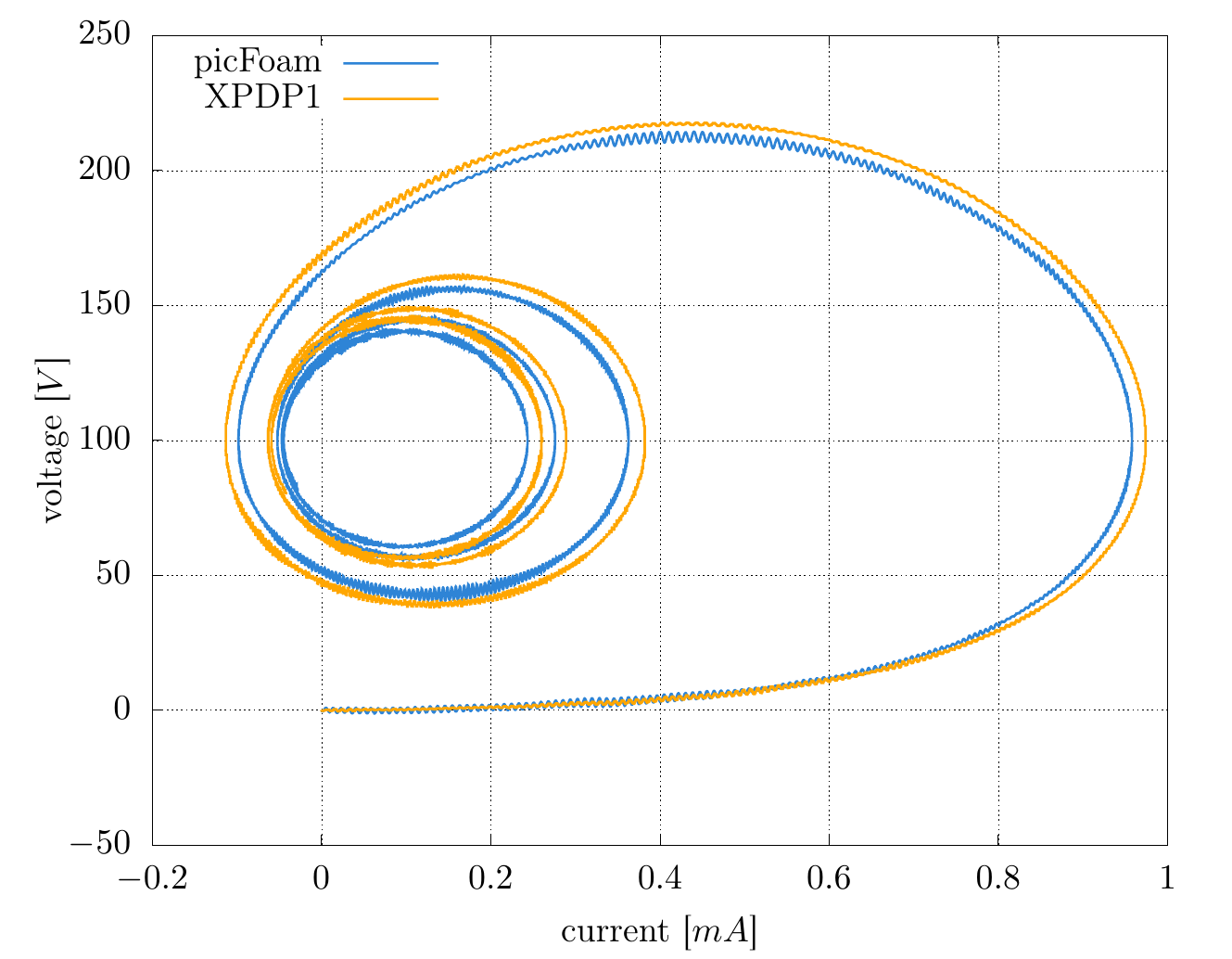}.
	\caption{Behavior of an RLC circuit with $\textnormal{R} = 1\,\Omega$, $\textnormal{L} = 0.04\,\textnormal{H}$ and $\textnormal{C} = 1\,\mu\textnormal{F}$ interacting with an argon plasma of density $n_e = n_i = 10^{15}\,\textnormal{m}^{-3}$; comparison between XPDP1 and picFoam.}
	\label{fig:RLCplasma}
\end{figure}

The second test looks at the evolution of current and voltage at the circuit driven boundary over time. Compared to the first validation, the 1D domain of $5\,\textnormal{mm} \times 2.4\,\textnormal{mm} \times 2.4\,\textnormal{mm}$ set up with 200 cells, is filled with an argon plasma of density $n_e = n_i = 10^{15}\,\textnormal{m}^{-3}$, where the subscript indicates the electron and ion species respectively. The species are initialized in thermal equilibrium using a Maxwell-Boltzmann distribution and a temperature of $\textnormal{T}=1$ eV. Collisions between species are disabled and the circuit elements are set to $\textnormal{R} = 1\,\Omega$, $\textnormal{L} = 0.04\,\textnormal{H}$ and $\textnormal{C} = 1\,\mu\textnormal{F}$. The solution of the plasma influenced circuit behavior is compared to a simulation with XPDP1\cite{Verboncoeur90} using the same parameter as in picFoam. Fig.\ref{fig:RLCplasma} shows the voltage plotted over the current for both codes, here showing good agreement between these two implementations with a mean error of 3\% for the measured current.

\section{Discussion}
\label{sec:Discussion}

In this work we have presented the implementation and validation of picFoam, a fully kinetic electrostatic solver implementing the Particle-in-Cell method including Monte Carlo Collisions for plasma research. The solver is developed in OpenFOAM building upon a robust particle tracking algorithm and the finite volume method. picFoam incorporates the same object-oriented design as the OpenFOAM toolbox, which makes it highly flexible by being able to select the different implemented models at run time and easy to extend because of the modular design, using C++ template programming techniques. 
picFoam implements state of the art models for all of its submodules, this includes particle collisions both relativistic and non-relativistic among particles whose weights are arbitrary and defined on a particle to particle basis. OpenFOAM's novel barycentric particle tracking algorithm enables picFoam to efficiently manage the linear interpolation of the charge and field weighting from and to meshes, on one to three spacial dimensions. Moreover, its application of general circuit boundary models allows picFoam to simulate real plasma devices.\\
Validations for all implemented submodules have been presented, demonstrating picFoam's ability to reproduce plasma phenomena. In section \ref{sec:ValidationPusher} and \ref{sec:PlasmaOscillation} we have shown fundamental validations for the PIC method. In these simulations we checked the integration of motion by comparing the motion of charged particles against theory, here validating the consensus of theory and simulation, since errors are far bellow 1\% for all conducted tests. The simulation of plasma oscillation has shown excellent results for the implemented volume weighting method, which in comparison to the cell average method introduces far less numerical noise to the simulation. By employing barycentric particle tracking a novel approach of charge and field weighting for mesh based PIC methods without computational expensive particle search algorithms has been presented. Simulations for the implemented collision algorithms have been shown in section \ref{sec:ValidationCollision}, all algorithms were validated independently, overall conforming their correct implementation. Simulations in this section were run using three different ratios of particle weights, this also validating the correction algorithms needed for the collision of arbitrarily weighted particles. Small deviations from theory, seen here during simulations, can be accounted to the assumptions of a Maxwell-Boltzmann distribution incorporated into the theories. Boundary models are validated in section \ref{sec:BoundaryValidation}, here results obtained by picFoam agree with theory as well as with results achieved with other solvers.\\
The intention of picFoam is to be a useful platform for plasma research, it is easy to set up new simulation cases by employing OpenFOAM's strait forward case structure, as well as easy to extend through its modular design.\\
The development of picFoam is an ongoing process adding new features and bug fixes to its open source repository. Future efforts in extending picFoam's features include adding load balancing and the coupling to OpenFOAM's powerful field solver capabilities. This coupling of fluid and Lagrangian description would allow picFoam the simulation of individual species as fluid and thereby gaining the ability of conducting higher performing hybrid simulations. Additionally, in the context of noise introduced by the weighting methods (see section \ref{sec:PlasmaOscillation}), the addition of filtering algorithm, as mentioned in Jacobs et al.\cite{Jacobs2006}, poses the opportunity for extending picFoam, thereby increasing the robustness of the code.

\section*{Acknowledgments}

We thank the DFG Deutsche Forschungsgemeinschaft (German Research Foundation) for funding this project under grant GR 2720/8-1. We also like to thank Will Bainbridge (CFD Direct) for his help in understanding OpenFOAM's barycentric particle tracking.





\bibliographystyle{elsarticle-num}
\bibliography{literature}







\end{document}